\documentclass[aps,floats,amssymb,amsmath,prb,twocolumn,superscriptaddress,showpacs]{revtex4}
\usepackage{calc}
\usepackage{graphicx}
\usepackage{color}

\begin{document}
\title{Finite temperature crossovers and the quantum Widom line near the Mott transition}
\author{J. Vu\v ci\v cevi\'c}
\affiliation{Scientific Computing Laboratory, Institute of Physics
Belgrade, University of Belgrade, Pregrevica 118, 11080 Belgrade,
Serbia}
\author{H. Terletska}
\affiliation{Condensed Matter Physics and Materials Science Department,
Brookhaven National Laboratory, Upton, New York 11973, USA}
\author{D. Tanaskovi\'c}
\affiliation{Scientific Computing Laboratory, Institute of Physics
Belgrade, University of Belgrade, Pregrevica 118, 11080 Belgrade,
Serbia}
\author{V. Dobrosavljevi\'{c}}
\affiliation{Department of Physics and National High Magnetic Field Laboratory,
Florida State University, Tallahassee, Florida 32306}

\begin{abstract}
{
The experimentally established phase diagram of the half-filled Hubbard model features the existence of three distinct finite-temperature regimes, separated by extended crossover regions. A number of crossover lines can be defined to span those regions, which we explore in quantitative detail within the framework of dynamical mean-field theory. Most significantly, the high temperature crossover between the bad metal and Mott-insulator regimes displays a number of phenomena marking the gradual development of the Mott insulating state. We discuss the quantum critical scaling behavior found in this regime, and propose methods to facilitate its possible experimental observation. We also introduce the concept of {\em quantum Widom lines} and present a detailed discussion that highlights its physical meaning when used in the context of quantum phase transitions.
}
\end{abstract}


\maketitle

\section{Introduction}

Strongly correlated materials exhibit a variety of phases whose
properties often lack a complete microscopic
understanding.\cite{Dobrosavljevic_book2012} The most interesting
new aspect of this class of materials is a possibility to tune the
system through two or more different ground states separated by
quantum critical points (QCPs).\cite{Sachdev_book} Such QCPs are
often difficult to directly approach and investigate, not only
because they reside at $T=0$, but also because various additional
instabilities and orders emerge in their immediate vicinity.
Nevertheless, understanding them is of chief importance, because
they often control rather extended finite temperature quantum
critical regions displaying universal properties and featuring
scaling behavior of all quantities.

Quantum critical points have been experimentally identified and studied in
several classes of physical systems, ranging from heavy fermion metals\cite{stewartrev1,Lohneysen2007} to conventional\cite{re:Sondhi97} and even
high temperature superconductors.\cite{christos-vlad2005prb} In most of these, however, the QCP is
obtained when quantum fluctuations become sufficiently strong to suppress
an appropriate ordering temperature - for magnetic, structural, or superconducting
order - down to $T=0$. When this happens, then concepts familiar from
the very successful theory of classical critical phenomena can be utilized
and naturally extended  to a quantum regime.\cite{Sachdev_book} Indeed,
most conventional  theoretical approaches follow the Landau theory
paradigm\cite{landau1957} and examine the impact of thermal and
quantum fluctuations  of appropriate order parameters, as describing the
corresponding patterns  of spontaneous symmetry breaking.

Should most exotic phenomena, then, be regarded as manifestations
of some form of (static or fluctuating) order, as Slater
speculated even in 1930s,\cite{slater34rmp} or should
fundamentally different classes of quantum phase transitions exist? The first viewpoint
was at the origin of the Hertz (weak coupling)
approach\cite{hertz-prb76,millis-prb93} to quantum criticality,
which despite its formal elegance resulted in only modest
successes. The latter, however, was at the core of pioneering
ideas of Mott\cite{mott1949}  and Anderson,\cite{andersonlocrev}
who provided a complementary perspective. According to their
views, strong electronic correlations are able destroy the
metallic state even in absence of any ordering, leading to the
formation of the Mott insulating state. The existence of broad
classes of Mott insulators is, of course, beyond the doubt at this
time. And while most order anti-ferromagnetically at low
temperature, they indeed remain robustly insulating (gaps often in
the electron-Volt range) even well above the corresponding Neel
temperature.\cite{Imada1998,Kanoda_review2011,Powell_review2011}

The nature of the phase transition between the metallic and the insulating phase - the
Mott transition - has, in contrast, remained highly controversial and subject to much
debate. Because the two phases share the same symmetries, the clear distinction
between them is apparent only at $T=0$. Should a direct and {\em continuous} transition
between a paramagnetic metal and a paramagnetic Mott insulator exist at $T=0$,
it would represent the most obvious example of a QCP outside the Landau paradigm -
unrelated to any mechanism of spontaneous symmetry breaking. Unfortunately, in most
familiar situations, the Mott metal-insulator transition is also accompanied by
simultaneous magnetic, charge, structural, or orbital ordering, considerably
complicating the situation and fogging the issues, both from the theoretical and
the experimental perspective.

Still, it is a well established experimental fact that in
all known cases, the characteristic temperature scale $T_c$,  below which many of such
"intervening" phases are found, is quite small - as compared to both basic
competing energy scales: the Fermi energy $E_F$ measuring the quantum fluctuations,
and the Coulomb repulsion $U$ that opposes the electron motion.
As a result, a very sharp crossover between metallic and insulating behavior is observed
even at $T \gg T_c$,  for all physical quantities.
The key issue thus remains: what is the main physical mechanism controlling this finite temperature
metal-insulator crossover? Should it be viewed as a quantum critical regime dominated by
appropriate order-parameter fluctuations, or is it - as postulated by Mott and Anderson -
a dynamical phenomenon not directly related to any ordering tendency.

To clearly and precisely address this question, one must: (1) Suppress all ordering tendencies,
at least in the relevant temperature range, and (2) Understand and describe the remaining
physical processes controlling the resulting finite temperature crossovers, and the
corresponding quantum critical region, if one exists. From the theoretical point of view, this ambitious
goal is generally very difficult to achieve, at least for realistic model systems. The task is hard,
because standard perturbative approaches, which are so well-suited to describe
Fermi surface instabilities and the associated
competing orders, are quite incapable in describing the Mott physics. The situation, however, improved
with the development of dynamical mean-field theory (DMFT) method,\cite{Georges1996} which
capitalizes on performing a local approximation for appropriate self-energies and vertex functions, yet which
provides a completely non-perturbative description of strong correlation effects. Its physical content is most clearly
revealed by focussing at the "maximally frustrated Hubbard model" (MFHM)\cite{Georges1996,Terletska2011} with
long range and frustrating inter-site hopping (see below), where the DMFT approximation becomes exact.

%\section{maximally frustrated Hubbard model}

The MFHM - because it is maximally frustrated - displays no magnetic or any other
kind of long range order across its phase diagram. It does display, however, a
precisely defined Mott metal-insulator transition at low temperature, precisely in the fashion
anticipated by early ideas of Mott and Anderson. It has been studied by many authors,
ever since the beginning of the DMFT era some twenty years ago\cite{re:Rozenberg92}; yet surprisingly, some
of its basic features have remained ill-understood and even confusing. Most studies focused
on characterizing the low temperature behavior, where a strongly correlated Fermi liquid (FL)
forms on the metallic side of the Mott transition.\cite{re:Rozenberg92} At low temperatures, this FL phase is separated
from the Mott insulator by an intervening phase coexistence region (see Fig. 1), and the associated
first-order transition line (FOTL) terminating at the critical end-point (CEP) at $T=T_c$.\cite{Kotliar2000} The behavior
in the immediate vicinity of the CEP has attracted much recent attention\cite{Limelette2003,Kagawa2005} but unsurprisingly (as any
other finite-temperature CEP), it display scaling behavior of the standard classical liquid-gas (Ising)
universality class.\cite{Kotliar2000} Indeed, several experiment reporting transport in this regime have successfully been
interpreted\cite{ising-resistors} using these classical models.

But what about the supercritical ($T \gg T_c$) behavior? Its rough features have been investigated
by many authors,\cite{Georges1996} who identified several regimes and complicated crossovers connected them,
but no simple and plausible physical picture has emerged. Most importantly, almost no one has
attempted to interpret the features of this high temperature regime in terms of ideas or concepts
of quantum criticality.\cite{Sachdev_book} The complication, of course, comes from the presence of the coexistence dome
at $T < T_c$, which confuses the issues, and - at least at first glance - makes the situation seem
incompatible with the standard paradigm of quantum criticality.

\begin{figure}[t]
\includegraphics  [width=3.2in]{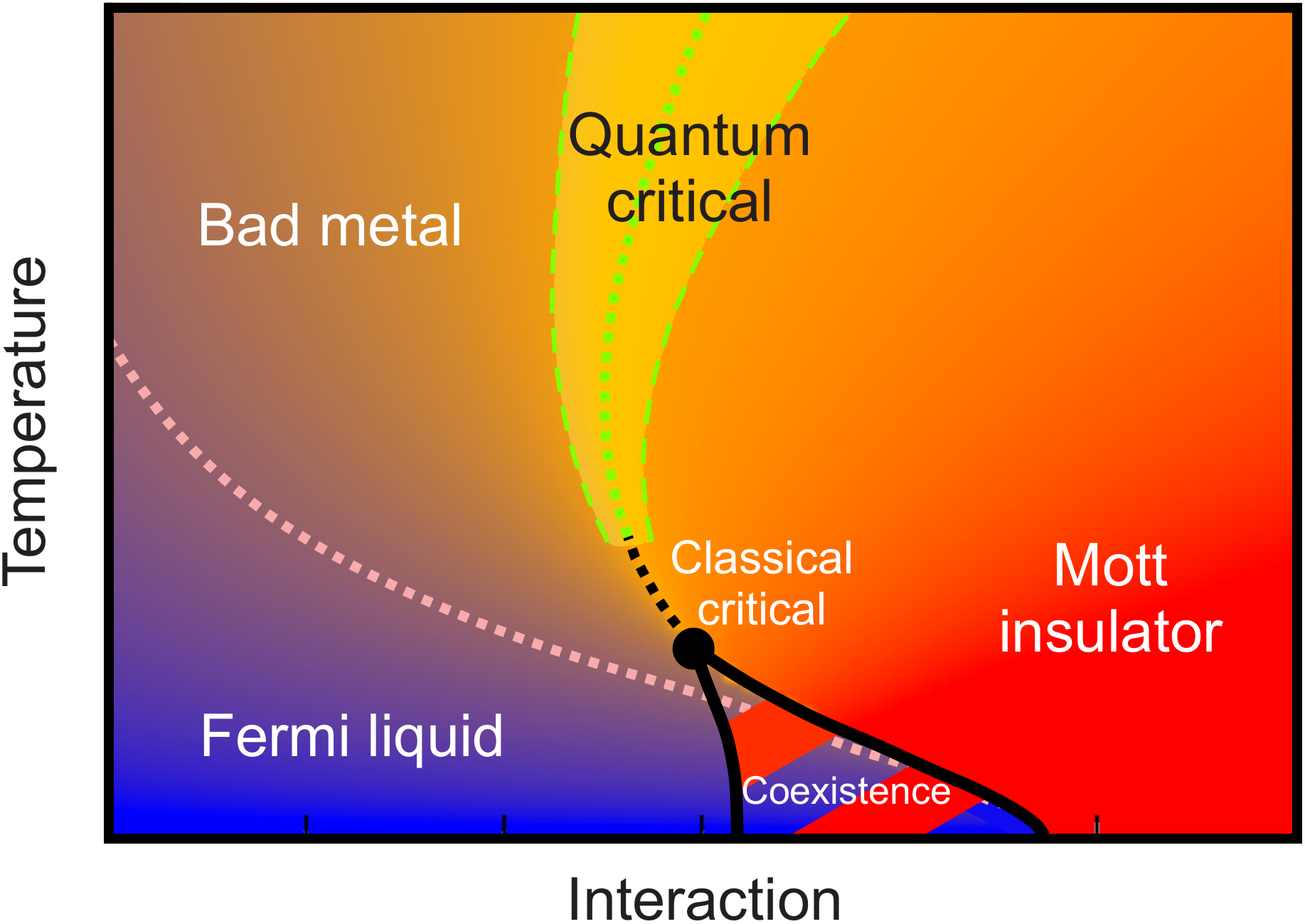}
\caption{Phase diagram of the half-filled maximally frustrated
Hubbard model. The background is an actual color map of the
resistivity obtained using the IPT impurity solver (see the text):
blue - small resistivity, red color - large resistivity. }
\label{phase_diagram}
\end{figure}

Our very recent work,\cite{Terletska2011} however, provided a new perspective. It made two key observations: (1) The characteristic
temperature scale of the coexistence dome $T_c \ll E_F, U$. The physics associated with it should,
at $T \gg T_c$, be little affected by its presence, and thus behave just as if $T_c \approx 0$, and
an actual QCP would exist separating the two phases; (2) To reveal the possible quantum critical scaling
associated with the proposed "hidden" QCP, one must follow a judiciously chosen trajectory (sometimes
called the "Widom line"\cite{Simeoni2010,Sordi2012}), as in almost any
standard critical phenomenon. This work also demonstrated\cite{Terletska2011}  remarkable scaling of the
resistivity curves, displaying all features expected of quantum criticality. The resistivity around this line
exhibits a characteristic ``fan-shaped'' form, surprisingly
similar to experimental findings in several systems,\cite{Dobrosavljevic_book2012,Abrahams2001,
Dobrosavljevic1997,Radonjic2012,Limelette2003,Kagawa2005}, reflecting gradual crossover from metallic to insulating
transport. The scaling behavior in this high-temperature crossover
regime was thus argued to encapsulate the universal
features of finite temperature transport near the metal-insulator transition.

The work of Ref.~\onlinecite{Terletska2011} focused on behavior close to the "instability line" and the associated
quantum critical scaling regime around it. It should be noted, however, that several other finite-temperature
crossover lines have been discussed by other authors\cite{Georges1996,Pruschke1993,Rozenberg1995,haule07hightc,Sordi2012}
to characterize the metal-insulator
region. The exact relationship between these different ideas and approaches - for the same model - thus remained
an open and rather confusing issue, that needs to be carefully investigated and understood. This important task
is the chief subject of this paper, where we present a detailed and very precise characterization of all the crossover
regimes across the entire phase diagram for the maximally frustrated Hubbard model at half-filling, within the
paramagnetic solution of dynamical mean field theory.
We carefully characterize the relevant crossover lines employing all the various proposed criteria used for their definitions. Two fundamentally distinct crossover regions are identified: one referring to the thermal destruction of long-lived quasiparticles and the other to the gradual opening of the Mott gap.
The instability line, as previously determined from a thermodynamic analysis,\cite{Terletska2011} belongs to the latter region, and is found to lie very near to the line of inflection points in the resistivity curves $\log \rho(U)$. The scaling of resistivity curves found around both of these lines is analyzed and discussed from the perspective of hidden quantum criticality and its experimental observation. In the end, we outline the generalized concept of the Widom lines,
and argue that they gain a new fundamental meaning in the context
of quantum phase transitions, which opens an avenue to put our results
into more general theoretical framework.

\section{Phase diagram}

We consider a single band Hubbard model at half-filling
\begin{equation} \label{hamiltonian}
H=-t\sum_{<i,j>\sigma}\left(c_{i\sigma}^{\dagger}c_{j\sigma}+h.c.\right)+U\sum_{i}n_{i\uparrow}n_{i\downarrow},
\end{equation}
where $c_{i\sigma}^{\dagger}$ and $c_{i\sigma}$ are the electron
creation and annihilation operators,
$n_{i\sigma}=c_{i\sigma}^{\dagger}c_{i\sigma}$, $t$ is the nearest-neighbor
hopping amplitude, and $U$ is the repulsion between two electrons
on the same site. We use a semicircular density of states, and the
corresponding half-bandwidth $D=2t$ is set to be our energy unit.
We focus on the paramagnetic DMFT solution, which is formally
exact in the limit of large coordination number, including the maximally-frustrated Hubbard model.\cite{Georges1996,Terletska2011} The DMFT provides a unique
theoretical framework, as it works well in the entire range of model parameters,
thus treating all the relevant phases and regimes on an equal footing. It is, however, most reliable at
high temperatures,\cite{Georges2011,Tanaskovic2011,Liebsch2009,Kokalj2012} when the correlations are more
local, and this is precisely the regime of primary interest of this paper. To solve the DMFT
equations we utilize both the iterated perturbation
theory{\cite{Georges1996}}  (IPT)  and the numerically exact
continuous time quantum Monte Carlo (CTQMC).\cite{Werner2006,Haule2007} The results obtained with these two methods are found to be in very good agreement. In this section we concentrate on IPT results, which cover the entire phase diagram and do not suffer from numerical noise. Figures in the rest of the paper are the QMC results.

The phase diagram in the $U-T$ plane is shown in Fig.~\ref{phase_diagram}.
The DMFT solution reproduces the three regimes found close to the metal-insulator transition (MIT): Fermi liquid,
bad metal and Mott insulator, in qualitative agreement with experiments on various Mott systems.\cite{Georges1996}
We begin their characterization by first analyzing the behavior of the resistivity in the relevant range of parameters.

The DMFT expression for the calculation of DC resistivity, $\rho=1/\sigma(\omega\rightarrow 0)$, is given by \cite{Georges1996}
\begin{equation} \label{conductivity}
\sigma=\pi \sigma_0 \int_{-\infty} ^{+\infty}d\varepsilon v^2
(\varepsilon)D^0(\varepsilon)\int_{-\infty} ^{+\infty}
(-\frac{df}{d\omega}A^2(\varepsilon,\omega)),
\end{equation}
where $A(\varepsilon,w)=-\frac {1}{\pi}ImG(\varepsilon,w)$,
$v(\varepsilon)=\sqrt{(4t^2-\varepsilon^2)/3}$.
$D^0(\varepsilon)=\frac{1}{2\pi t^2}\sqrt{4t^2-\varepsilon^2}$ is
the noninteracting DOS, and $f$ is the Fermi function.
The calculation of resistivity from the IPT results is straightforward as this method is defined on the real axis.
To calculate the resistivity from the QMC results, one first needs to perform the analytical continuation, which we carry out using the maximum entropy method.\cite{Jarrell1996}

\begin{figure}[h]
\includegraphics  [width=3.2in]{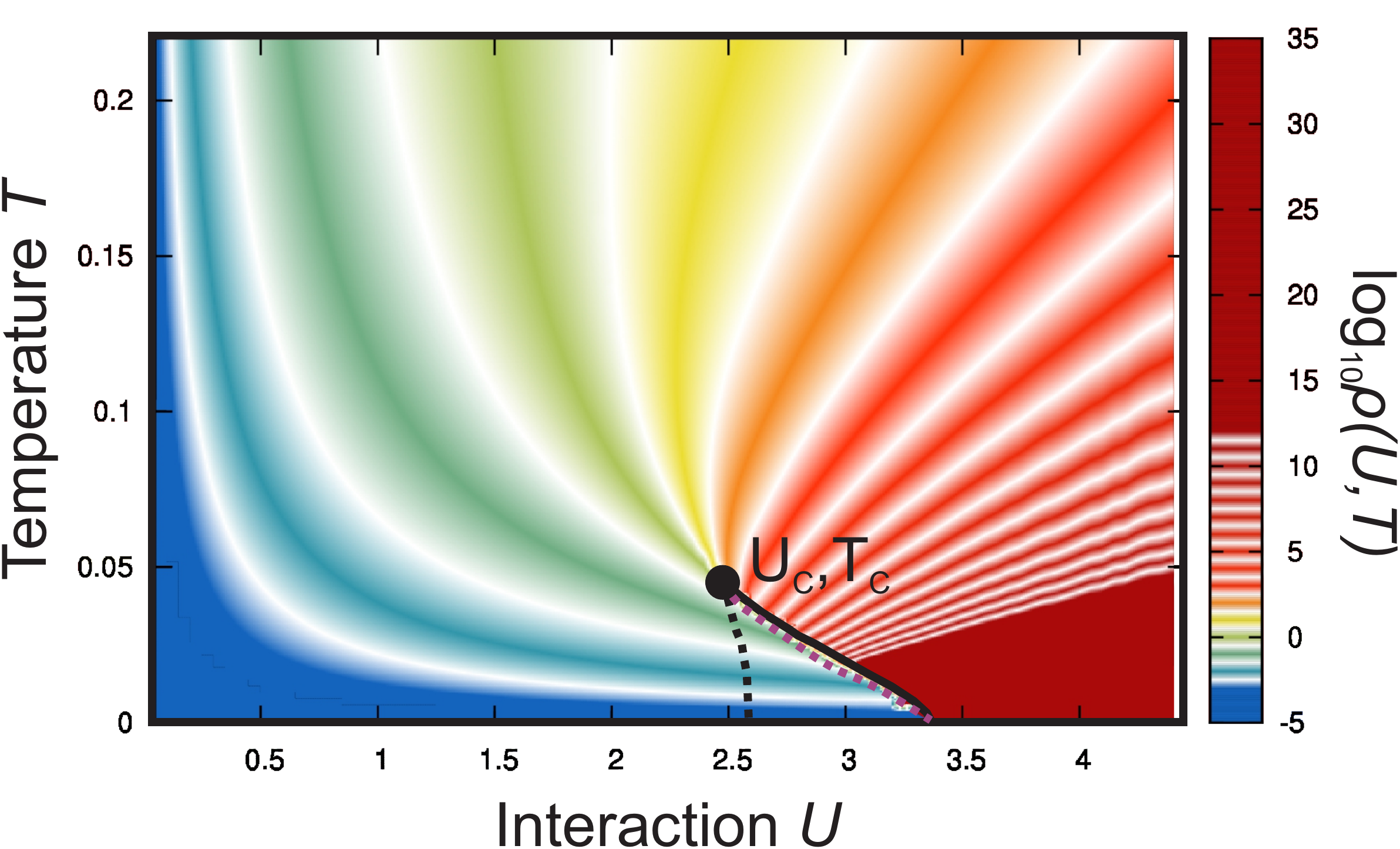}
\caption{Resistivity (in units of $\rho_{_{Mott}}$) calculated in the entire $U-T$ plane. The white
stripes follow the lines of equal resistivity and separate the
orders of magnitude in the resistivity. Spinodals are denoted with thick black lines, and
the first order phase transition line is dashed. } \label{rho}
\end{figure}

Our quantitative IPT results are replotted in Fig.~\ref{rho}, where the value of resistivity is color-coded, with white stripes
separating the consecutive orders of magnitude between $10^{-3}$ and $10^{13}$.
In this plot, as well as in the rest
of the paper, the resistivity is given in the units of $\rho_{_{Mott}}$, the maximal metallic resistivity
in the semiclassical Boltzmann theory, defined as the resistivity of the
system when the scattering length is equal to one lattice spacing.\cite{Hussey2004,Merino2000}
At zero temperature, the metallic resistivity vanishes, while the Mott insulator has an infinite resistivity.
With increasing temperature, the difference between the two states becomes less and less pronounced.
[Between the spinodals, both metallic and insulating solutions are possible, but in this plot only the metallic resistivity is shown.] In the intermediate correlation, $U < U_c$, high temperature, $T > T_c$ regime, the resistivity is comparable or even larger then $\rho_{_{Mott}}$, but it still (weakly) increases with temperature, which is characteristic for the "bad metal" regime observed in several Mott systems.\cite{Hussey2004}

It is remarkable how this way of presenting the data immediately creates the familiar "fan-shape" structure, generally expected for quantum criticality.\cite{Sachdev_book} At high temperatures all the white constant-resistivity stripes seem to converge almost to the same point $U \sim U_c$. The perfect convergence, however, is interrupted by the emergence of the coexistence done at $T < T_c$, but such behavior is exactly what one expects for "avoided quantum criticality",\cite{haule07hightc} consistent with the physical picture proposed in Ref.~\onlinecite{Terletska2011}.

\begin{figure}[h]
\includegraphics  [width=3.2in]{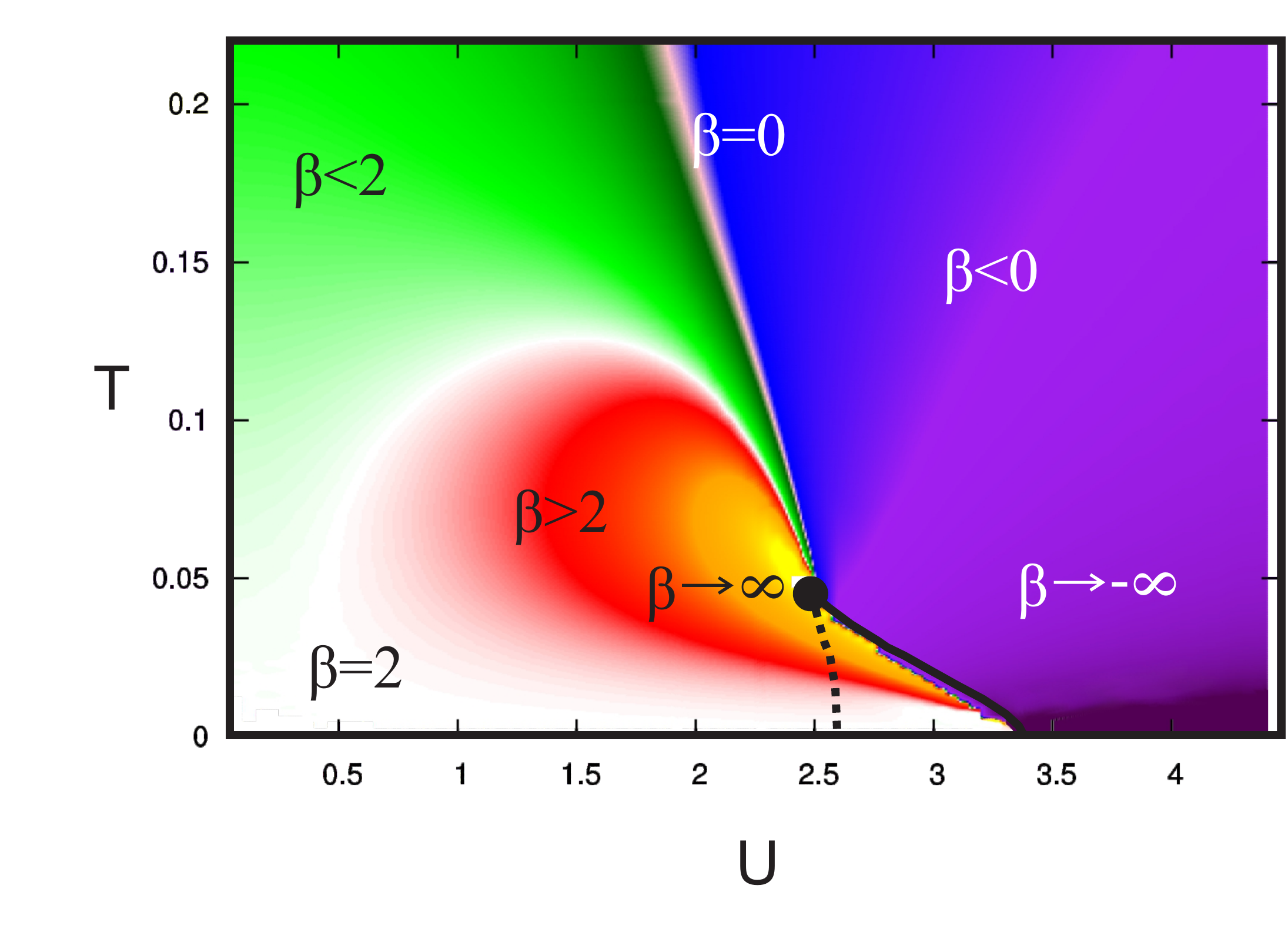}
\caption{The effective resistivity exponent ($\beta = d\log \rho / d\log T$) calculated in the
entire $U-T$ plane illustrates the different transport regimes (see the text).
}
\label{beta}
\end{figure}

Different regions of the phase diagram are also distinguished by the qualitatively different form for the temperature dependence of the resistivity. To make this behavior even more apparent,  we follow a commonly-used procedure to displaying the data around QCPs,  compute the logarithmic derivative of resistivity
with respect to the temperature, i.e. the "effective exponent"\cite{Cooper2009,Gegenwart2008}
\begin{equation}\label{beta_eq}
\beta (T,U) = d\log \rho (U,T) / d\log T,
\end{equation}
which is presented in color-coded form in Fig.~\ref{beta}.

On the metallic side, at the lowest temperatures, one finds a typical metallic dependence
of the form $\rho \sim T^2$ and here we have $\beta = 2$ (white). Far from the transition,
this regime survives up to relatively high temperatures, but eventually
the temperature dependence of the resistivity starts gradually slowing down, displaying behavior sometimes described as "marginal Fermi-liquid" transport  (green, $\beta \sim 1$). Closer to the transition,
this is preceded by an increase in the effective exponent (red), which is a reflection
of the existence of the critical end-point in which $\beta$ diverges (yellow).
Very close to the transition, a maximum
of the resistivity is reached at some temperature (pink) and the trend of the resistivity increase is then reversed.
On the other side of the phase diagram, deep in the Mott insulator,
one finds typical activation curves which
exhibit the exponential drop in the resistivity with increasing temperature,
due to the gap in the excitation spectrum (black and purple). However, just above the coexistence dome,
one finds an intermediate regime, where the behavior is generally insulating because
the resistivity decreases with temperature, but the gap is not yet fully open, and the
temperature dependence deviates from exponential (blue). This region is sometimes referred to as the "bad insulator."

\section{Crossover lines}

In the previous section we have characterized the different regimes in the vicinity of the Mott MIT:
Fermi liquid, bad metal and Mott insulator. However, apart from the coexistence region,
the properties of the system change continuously in the entire phase diagram.
The lines separating the different regimes are thus a matter of convention and many definitions can be found in literature proposing the criteria for their distinction.

\begin{figure}[t]
\includegraphics  [width=3.5in]{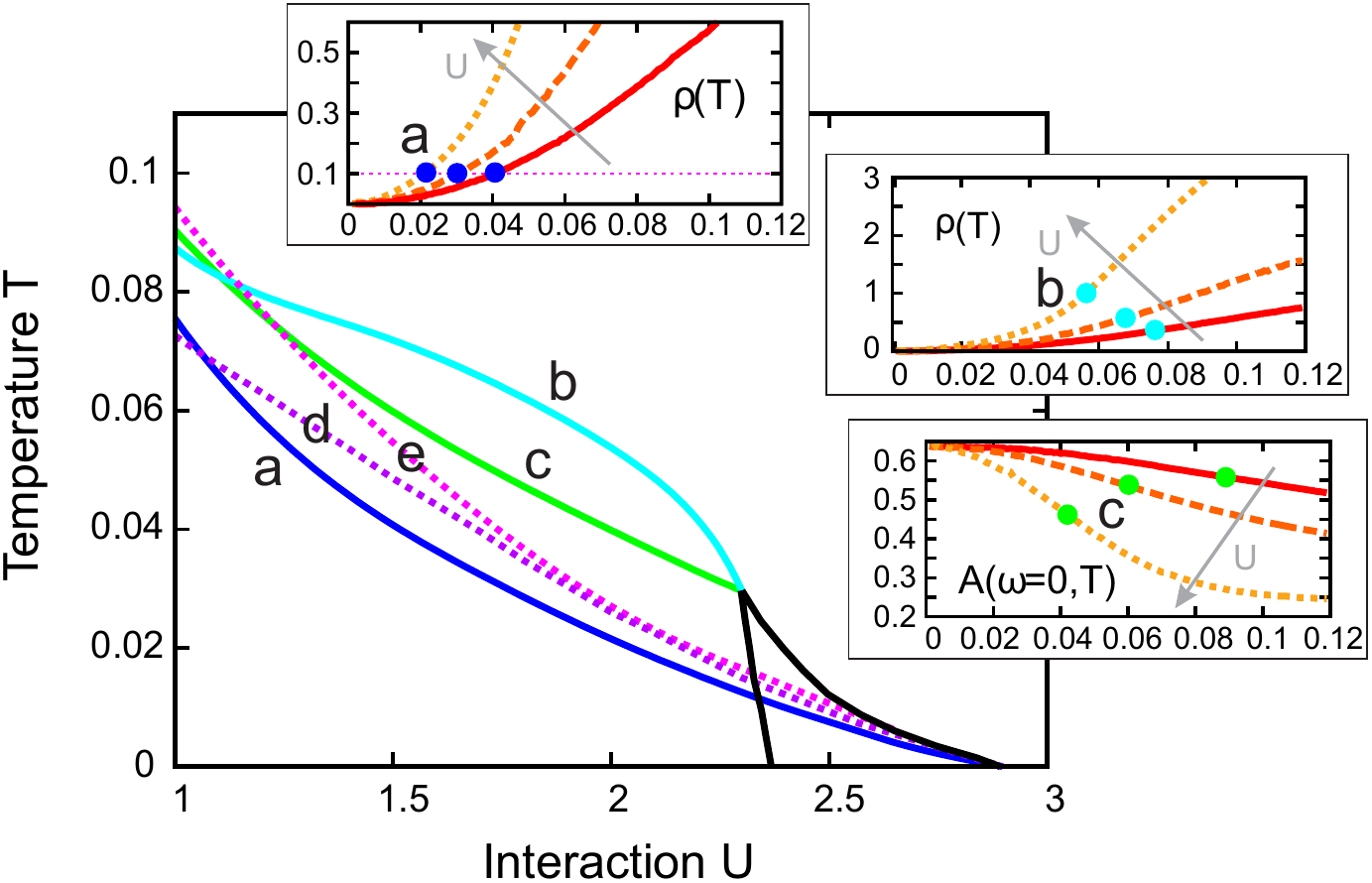}
\caption{ Various definitions for the crossover lines between
the Fermi liquid and the bad metal. The meaning of each definition is illustrated on a smaller panel to the right. The results are obtained with the QMC.}
\label{CrossoverLines_a}
\end{figure}

In Fig.~\ref{CrossoverLines_a} we present the lines corresponding to various definitions of a crossover line between the Fermi liquid and the bad metal regimes.
The definition of each line is illustrated on a smaller panel on the right, where the corresponding
feature in the resistivity and other relevant quantities is marked with the dots of the same color.
The dark blue line (a) is defined by
$\rho = 0.1 \rho_{_{Mott}}$ and it roughly corresponds
to the Fermi coherence temperature $T_{FL}$ (the temperature above which
the temperature-dependence of resistivity is no longer quadratic).
The corresponding small panel (a) shows the resistivity as a function of temperature, plotted for three different values of U.
The dotted horizontal line marks $\rho = 0.1 \rho_{_{Mott}}$. The arrow denotes the direction of increase of $U$.
The light blue line (b)
corresponds to the inflection point of the resistivity,
$d^2\rho(\omega=0)/dT^2=0$, and the green line (c) is determined as the inflection point of the
spectral density at the Fermi level with respect to the
temperature, $d^2A(\omega=0)/dT^2=0$. These are illustrated on smaller panels (b) and (c)
where the DC resistivity and $A(\omega=0)$ are plotted versus the temperature, for three different values of U.
The inflection points are marked with the dots of color corresponding to the (b) and (c) lines on the main panel.
The additional two dotted lines are: (d) the quasi-particle weight
at zero temperature defined by
$Z = \left[ 1 - \left. {d\mathrm{Im}\Sigma(i\omega_n)} /
{d\omega_n}\right|_{\omega_n \rightarrow 0} \right]^{-1}$
and (e) the zero temperature local spin susceptibility $\chi$. Both
quantities are divided by $10$ to fit in the temperature range of
the plot and to be more easily compared to the crossover lines. It
is evident that the coherence temperature is roughly proportional
to the quasi-particle weight at zero temperature, but with the
prefactor $~0.1$, $T_{\mathrm{FL}}(U) \sim 0.1\; Z(U)$. As compared with the doped Hubbard model,\cite{Deng2012,Xu2013} $T_{\mathrm{FL}}$ is higher, but still distinct from the temperature corresponding to $\rho_{_{Mott}}$, in agreement with the experiments on organic materials.\cite{Merino2000,LimelettePRL2003,Merino2008} The
quasiparticle weight $Z$ is weakly temperature dependent and the Drude peak in the opticalal conductivity is still pronounced for $\rho \lesssim \rho_{_{Mott}}$.\cite{Radonjic2010}

\begin{figure}[t]
\includegraphics  [width=3.5in]{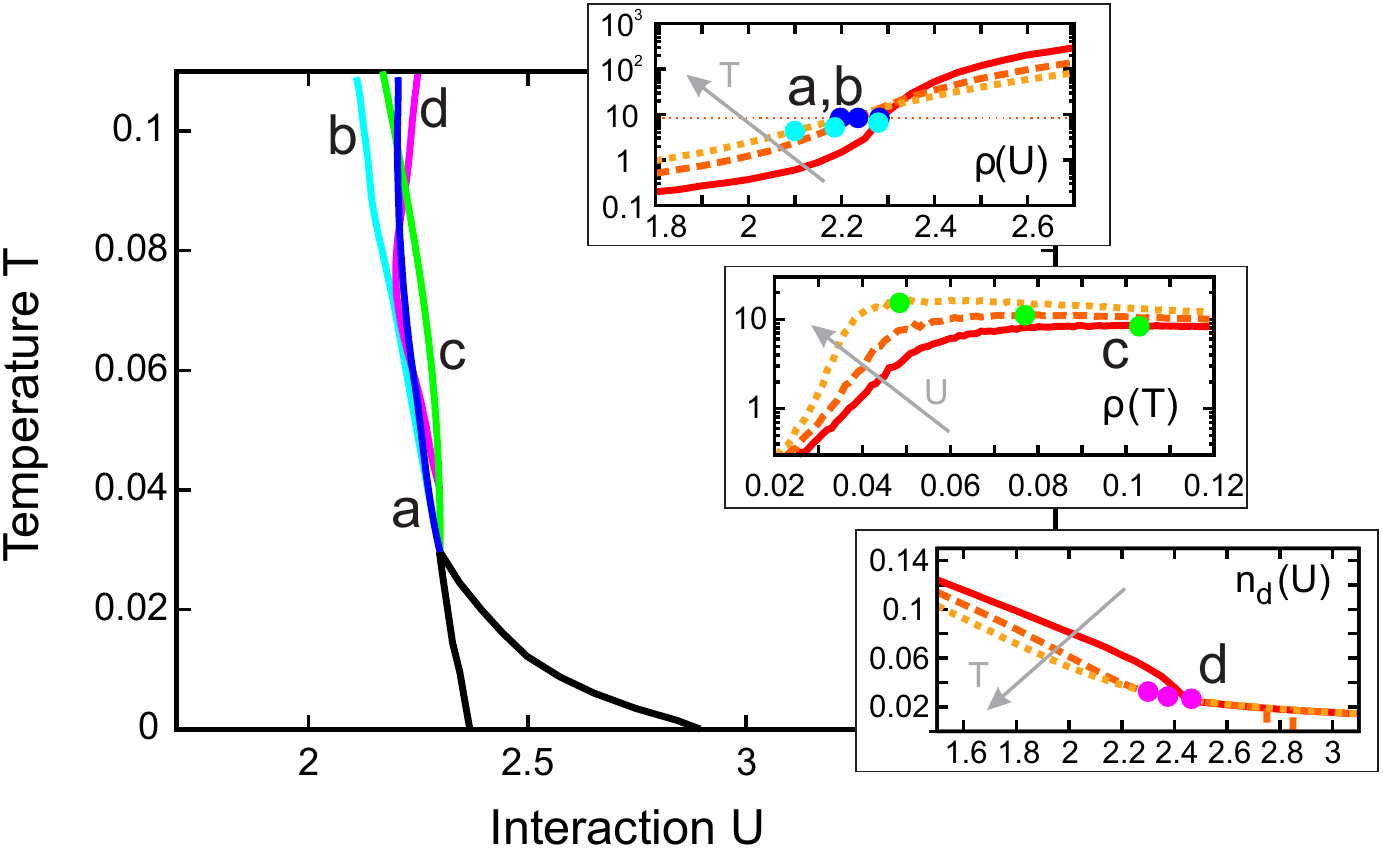}
\caption{ Various definitions for the crossover lines between the bad metal and the Mott insulator.
The meaning of each definition is illustrated on a smaller panel to the right. The results are
obtained with the QMC.}
\label{CrossoverLines_b}
\end{figure}

In contrast with these lines, one can also define the lines
separating the bad metal from the (bad) Mott insulator. In Fig.~\ref{CrossoverLines_b}, we present
several criteria for their definition.
In analogy to the line (a) of Fig.~\ref{CrossoverLines_a},
one can use the resistivity to distinguish between the two regimes.
The dark blue line (a) plotted here, connects the points where the resistivity is equal
to the one found precisely at the critical end-point, which we estimate to be roughly $10\rho_{_{Mott}}$.
The light-blue line (b) marks the inflection
point of logarithmic resistivity as a function of $U$ ($\partial^2 \log\rho(U,T)/\partial U^2=0$).
It is well pronounced feature up to high temperatures, and it is a direct consequence of the discontinuity
across the FOTL at $T < T_c$.
These two are illustrated on the small panel to the right, where $\log\rho(U)$ is plotted at three different temperatures.
The dark blue dots are the intersections of these lines with the dotted, $10\rho_{_{Mott}}$ line. The inflection points
are marked with the light blue dots, and are found at slightly lower values of U.
Another natural definition for the crossover is the $\beta=0$ line (c), as
it marks the place where the trend of resistivity growth is reversed.
At its right-hand side, the resistivity decreases with
temperature, which is a sign of insulating behavior.
This is illustrated on the corresponding small panel, where $\log\rho(T)$ is plotted for 3 different values of U and
the maxima are marked with the green dots.
The double occupancy $n_d$ has an obvious change in trend on
crossing the line (d). Here, the second derivative
$\partial^2  n_d / \partial U^2$ has a sharp maximum,
and separates the two distinct regimes of $ n_d(U)$, both almost linear but with different slopes.
This is apparent on the small panel (d) where double occupancy is plotted as a function of U at various temperatures.

It is striking that these lines almost coincide, in sharp contrast
to what is seen in Fig.~\ref{CrossoverLines_a}.
Altough the opening of the gap is very gradual, it is possible to
pin-point the boundary between the two regimes and actually divide
the supercritical part of the phase diagram into metallic and
insulating-like regions. In the following section we overview the
instability line, another definition for a metal-insulator
crossover line, and explain how it helps reveal a very peculiar
property of the Hubbard model, very suggestive when it comes to
interpreting the Mott MIT in terms of quantum phase transitions.

\section{Instability line and quantum critical scaling}

\begin{figure}[b]
\includegraphics[width=3.0in]{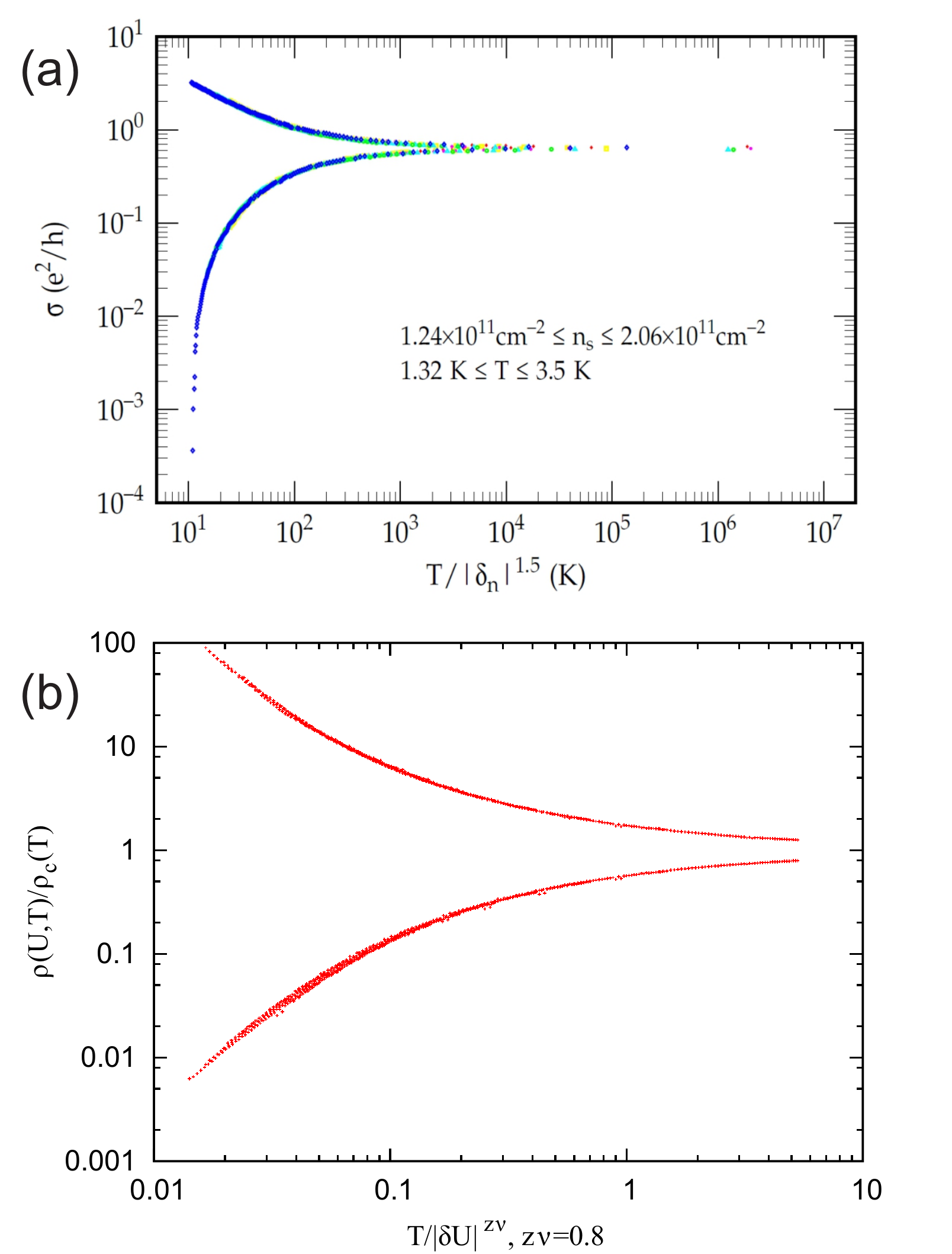}
\caption{(a) Experimental results: conductivity scaling in
high-mobility Si-MOSFETs presents a textbook example of the
quantum critical scaling (taken from Ref.~\onlinecite{Popovic1997}). (b) DMFT-QMC results:
resistivity scaling strongly reminiscent of what is seen in
MOSFETs. After dividing $\rho(U,T)$ with the value of resistivity
on the instability line $\rho_c(T)$ (see the text) and then rescaling the
temperature with an appropriately chosen parameter $T_0(\delta
U)$, the resistivity curves collapse onto two branches.} \label{Scaling}
\end{figure}

It is a well established phenomenon that in the vicinity of quantum critical points, at finite temperatures, physical observables display a characteristic quantum critical scaling.\cite{Sachdev_book} A very good example of this is the transport in high mobility two-dimensional electron gases, in particular in metal-oxide-semiconductor field-effects transistors (MOSFETs).\cite{Dobrosavljevic_book2012} There are overwhelming evidence that they exhibit a zero temperature metal-insulator transition
at a critical concentration of charge carriers.\cite{Abrahams2001} It is experimentally observed in these systems that the value of resistivity at finite temperatures above the quantum critical point $(n_c,T=0)$ is a function of only $\delta n=n-n_c$ and $T$, which is considered a hallmark of quantum criticality. As shown in Fig.~\ref{Scaling}(a),\cite{Popovic1997} the resistivity curves collapse onto two branches: the resistivity is first divided by the ``separatrix'' $\rho_c(T)=\rho(n_c,T)$ which weakly depends on the temperature, and then the temperature is scaled by $T_o(\delta n)=|\delta n|^{\nu z}$, yielding
\begin{equation} \label{mosfet_scaling}
\rho(\delta n, T) = \rho_c(T)f(\delta nT^{-1/\nu z}).
\end{equation}
The mechanism behind the physical picture of MOSFETs is still elusive,\cite{Radonjic2012} but similar is seen is various spin systems, where the physics is well understood.\cite{Sachdev_book}  When there is a well defined order parameter, the separatrix corresponds to the line of zero symmetry-breaking field, which is trivially a straight vertical line emanating from the quantum critical point.

Although our model does feature a FOTL, the critical temperature is actually very low ($T_c \approx 0.03$), which makes it reasonable to pursue a description of its supercritical region from the perspective of quantum criticality. This is the approach that we have taken in a recent work\cite{Terletska2011} where we have shown that in the Hubbard model, a quantum critical scaling of the resistivity curves does indeed hold (Fig.~\ref{Scaling}(b)).
There is an obvious analogy between the interaction $U$ in our model and the carrier density $n$ in MOSFETs, but it was not immediately clear what line $U_c(T)$ should correspond to the separatrix in our model. The phase transition in the Hubbard model does not break any symmetries and the first-order transition line is curved, which indicated that $U_c$ has possibly a non-trivial temperature dependence.

\subsection{The instability line}

Starting from the thermodynamic arguments,\cite{Kotliar2000,Mooij1973} we have defined the instability line $U^*(T)$  as the line which corresponds to the minimum curvature of the free energy functional ${\cal F}[G(i\omega_n)]$ with respect to $U$.\cite{Moeller1999} Above $T_c$ the system has a unique ground state which corresponds to the minimum of ${\cal F}[G(i\omega_n)]$. In this minimum, the curvature of ${\cal F}[G(i\omega_n)]$  is determined by the lowest eigenvalue $\lambda$ of the fluctuation matrix
\begin{equation}\label{fluctuation_matrix}
M_{mn} = \frac{1}{2Tt^2} \left. \frac{\partial^2 {\cal F}[G]}{\partial G(i\omega_m) \partial G(i\omega_n)}
\right|_{G=G_{DMFT}},
\end{equation}
where  $\delta G(i\omega_n) \equiv G(i\omega_n)-G_{DMFT}(i\omega_n)$, and $G_{DMFT}$ is the self-consistent solution of the DMFT equations. As explained in detail
in Supplementary notes of Ref.~\onlinecite{Terletska2011}, $\lambda$ can be obtained by monitoring the rate of convergence in the DMFT iteration loop. Close to the self-consistent solution, the difference between the consecutive solutions drops exponentially, with an exponent proportional to $\lambda$. We have
\begin{equation}
{\bf G}^{(n+1)}-{\bf G}^{(n)} = \delta {\bf G}^{(n)}=e^{-n\lambda} {\bf G}_{\lambda}(i\omega_n),
\end{equation}
where ${\bf G}_{\lambda}$ is the eigenvector of $\hat M$ corresponding to its lowest eigenvalue $\lambda$.

The curvature $\lambda$ is actually a very general quantity that describes the
response of the system to an infinitesimal external perturbation, which may be a time-dependent field of an arbitrary form.
As such, $\lambda$ is very important in describing a thermodynamical state close to the Mott MIT, since it has a fundamentally dynamic nature.
Indeed, $\lambda$ vanishes precisely at the critical end-point,
as the free-energy functional becomes flat around $G_{DMFT}$. This is directly connected to the critical slowing down of dynamics, which
manifests as the vanishing of a characteristic frequency scale.
Above $T_c$, $\lambda$ is related to the local
stability of a given thermodynamic state and has a minimum precisely where the system is the least
stable, or where its proximity to either competing phase is
equal. Therefore, the instability line which connects the minima of $lambda$ vs. $U$ is the closest analogy to the lines of zero symmetry-breaking field in systems with an order parameter.

\begin{figure}[t]
\includegraphics  [width=3.2in]{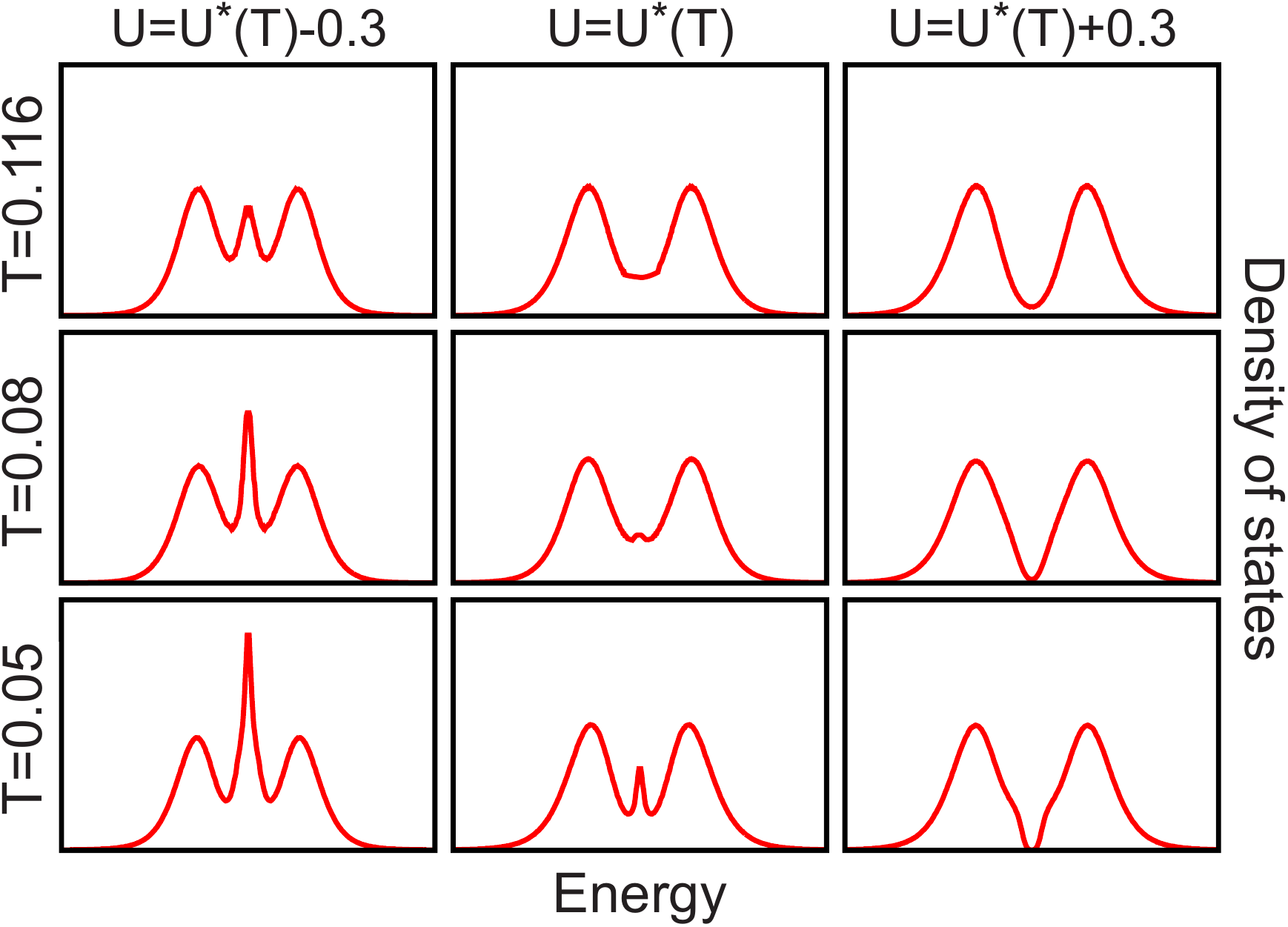}
\caption{ Density of states (QMC results) along the instability line $U^*(T)$
(middle column), and along the parallel trajectory for smaller
(left column) and larger $U$ (right column). } \label{WL_DOS}
\end{figure}

The instability line is presented in Fig.~\ref{phase_diagram} and indeed it represents a boundary between a metallic and insulating transport. It lies among the other crossover lines from Fig.~\ref{CrossoverLines_b}, (see also Section V). Its physical meaning is illustrated in Fig.~\ref{WL_DOS}
The middle column shows the DOS along the instability line for three different temperatures. While the DOS at the Fermi level is strongly suppressed, the gap is not yet fully open. Left column shows the density of states in the metallic phase following a trajectory parallel to the instability line: there is a clear quasiparticle peak at low temperatures, which gradually disappears as the bad metal region is reached by increasing the temperature. At larger $U$ (right column) the system is in the insulating phase with fully open Mott gap, featuring activated transport.

%------------------------------------------

\subsection{Free energy calculation}

To further illustrate the physical meaning of the instability line, we explore the free energy 
landscape in the Hilbert space of Green's functions.
For this we closely follow the procedure described in Ref.~\onlinecite{Moeller1999}.
The iterative self-consistency procedure used to solve the DMFT equations converges 
towards a local minimum of the corresponding “Ginsburg-Landau” free energy functional ${\cal F}[\bf{G}]$, which in the Hilbert space of the Matsubara Green’s functions $G(i\omega_n)$ takes the form

\begin{eqnarray} \label{free_energy}
{\cal F}[{\bf G}]&=&{\cal F}_{imp}[{\bf G}]+ {\cal F}_{bath}[{\bf G}], \\ \nonumber
    &=&{\cal F}_{imp}[{\bf G}]-t^2T \sum_{n} G^2(i\omega_n),
\end{eqnarray}
where the first term is the free energy of the impurity site in the
presence of the Weiss field ${\mathbf \Delta} = t^2 {\mathbf G}$, while the second term is  
is the energy cost of forming the Weiss field around a given site.

The DMFT self-consistency condition, typically reached via iterative procedure,
is then regarded as a saddle-point equation derived from the extremum condition of such Ginsburg-Landau
functional. The physical DMFT solution corresponds to the local stationary point of ${\cal F}[{\bf G}]$, 
where a gradient vector ${\bf g} = \partial {\cal F}[{\bf G}] / \partial {\bf G}$ becomes zero.
However, in the coexistence region below $T_c$, two such local minima are found. They correspond to physical solutions
(metallic ${\bf G}_M$ and insulating ${\bf G}_I$), and are separated by an unstable
solution (a local maximum or a saddle point). 

\begin{figure}[t]
\includegraphics[width=3.0in]{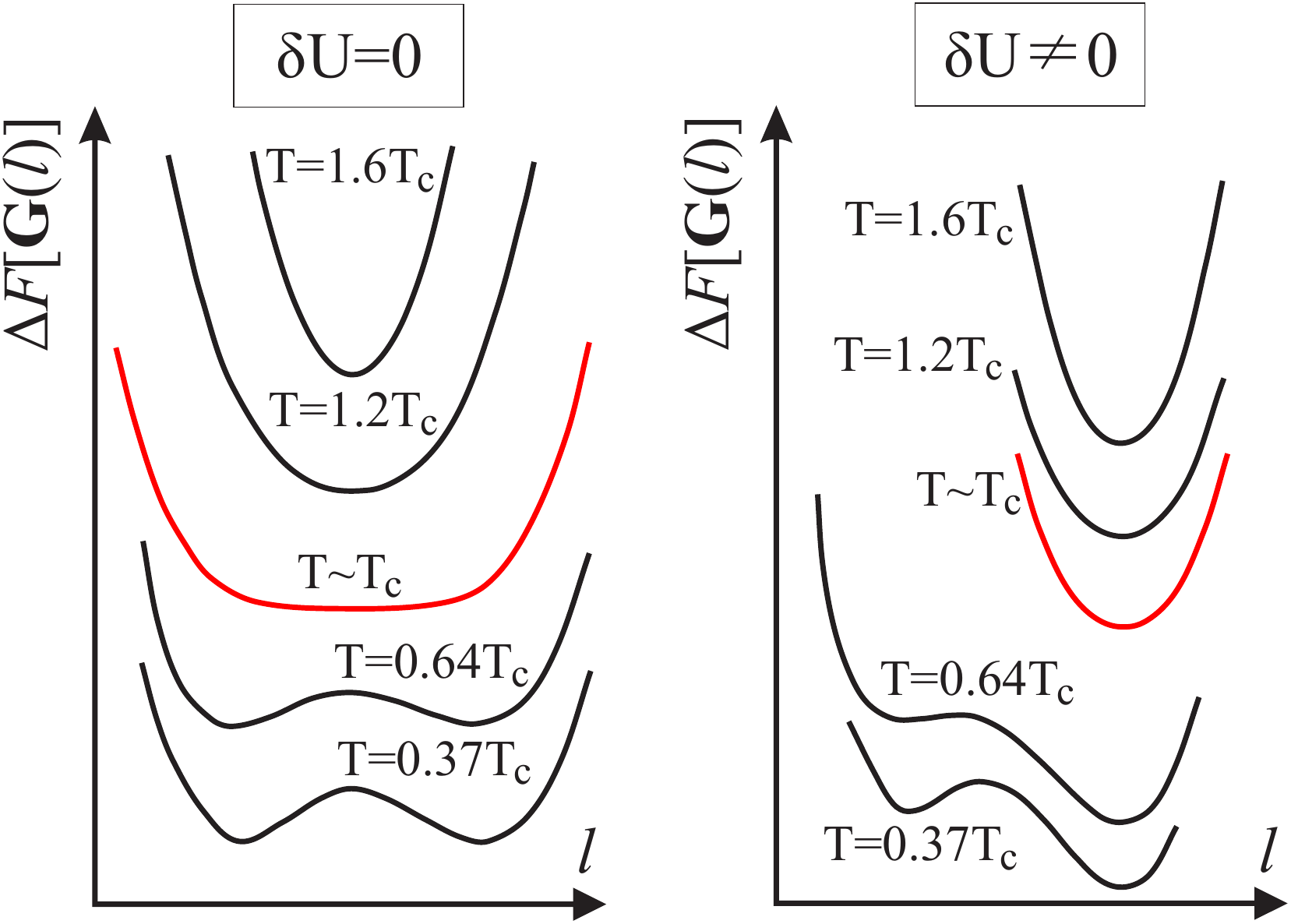}

\caption{Free energy landscape (IPT results):
(a) Along the ``zero field'' line ($\delta U=0$). At $T>T_{c}$,
the curvature of the free energy increases
with temperature, and it is zero at $T=T_{c}$. Below $T_{c}$, at the
first order transition line metallic and insulating solutions have the
same free energy. (b) Along the ``finite field'' line
($\delta U=-0.05$). At $T>T_{c}$, the curvature of the free energy
is greater than in the ``zero field'' case.
In the coexistence region one of the minima is energetically favored.
Note that the spacing between $\Delta {\cal F}$ curves for different temperatures is arbitrary.}
\label{FreeEnergy}
\end{figure}

We can visualize the shape of the infinitely dimensional free energy surface by calculating $F[\bf{G}]$ along a single direction going
through the self-consistent ${\bf G}_{DMFT}$. Below $T_c$, we do this along the direction connecting the two solutions, 
which can be parameterized as ${\bf G}(l)=(1-l){\bf G}_M-l{\bf G}_I$.
Above $T_c$, where there is only one solution, we follow the 
eigenvector ${\bf G}_{\lambda}$ with ${\bf G}(l)={\bf G}_{DMFT}+l{\bf G}_{\lambda}$. 
The relative change of the free energy is calculated\cite{Moeller1999} as an integral 
$\Delta {\cal F}(l)={\cal F}({\bf G}(l))-{\cal F}({\bf G}_{M/DMFT})=t^2T\int_0^l d{l'} {\bf e}_l \cdot {{\bf g}({\bf G}(l'))}$, 
where ${\bf e}_l$ is the unit vector of the followed direction (${\bf e}_l=({\bf G}_M-{\bf G}_I)/|{\bf G}_M-{\bf G}_I|$ below $T_c$ and 
${\bf e}_l={\bf G}_\lambda/|{\bf G}_\lambda|$ above $T_c$).
The gradient vector takes the form
${\bf g}={\bf G}_{imp}({\bf G})-{\bf G}$, with ${\bf G}_{imp}({\bf G})$ is the output of the impurity solver used in the DMFT procedure,
and $\bf{G}$ is the input - effective medium (hybridization bath) Green’s function.

Panel (a) shows the free energy landscape around ${\bf G}_{DMFT}$, precisely at the instability line. 
The curvature of the global minimum vanishes as one approaches $T_c$, which is consistent with eigenvalue $\lambda$ being zero at this point.
Below $T_c$ there are two minima and the instability line is no longer well defined, but it is logically continued to the line of the first order phase transition,
where two possible solutions are of the same energy. On panel (b), we move along a parallel trajectory, defined by $\delta U\neq 0$. 
It is immediately obvious that $\lambda$ never reaches zero and that in the coexistence region one of the solutions is energetically favoured.
This physical picture is common to various models. For example, it is seen in the Ising model in an external field, where the analogy is between the strength of the magnetic field and $\delta U$ in our case.

\subsection{Quantum critical scaling}

While the instability line is determined from the free energy analysis, a novel physical perspective is obtained by looking at the transport properties in its vicinity. We have demonstrated\cite{Terletska2011} that around this line, all resistivity curves can be collapsed onto two branches: we first divide each resistivity
curve by the resistivity along the instability line (the ``separatrix'') $\rho_c(T)=\rho(T,\delta U=0)$, and then rescale the temperature for each curve with an appropriately chosen parameter
$T_{0}(\delta U)$ to collapse the data onto two branches (Fig.~\ref{Scaling}(b)). The family of resistivity curves displays characteristic quantum critical scaling of the form
\begin{equation}\label{scaling_eq}
\rho(T,\delta U)=\rho_c(T)f(T/T_{o}(\delta U)),
\end{equation}
with $T_{o}(\delta U)\sim |\delta U|^{z\nu}$.
The scaling parameter $T_{o}$ displays power law scaling with the
same exponents for both scaling branches and falls sharply as $U
\rightarrow U^*$, which is consistent with quantum critical scenario. The resistivity scaling holds in the temperature range roughly between 2 and 4 $T_c$, as depicted in Fig.~\ref{phase_diagram}. We estimate the exponent $z\nu$ to be around $0.6$ when IPT is used to solve the DMFT equations. The scaling procedure with the data obtained with the CTQMC impurity solver gives slightly larger critical exponent with an error bar due to numerical noise of the data and due to the analytical continuation.

We emphasize the difference of the proposed quantum critical scaling and classical scaling in the immediate vicinity of the critical end-point (classical critical region in Fig.~\ref{phase_diagram}). It has been already carefully studied theoretically,\cite{Kotliar2000,Semon2012} and even observed in experiments, \cite{Limelette2003} revealing the classical Ising scaling in this regime. In contrast, the scaling parameter in our formula is $T$ rather than $|T-T_c|$ and the value of the exponent does not fit any of the known universality classes. The scaling region in our analysis is significantly broader and the collapse of the resistivity curves is observed in a large temperature region above the critical end-point.

\begin{figure}[t]
\includegraphics  [width=3.2in]{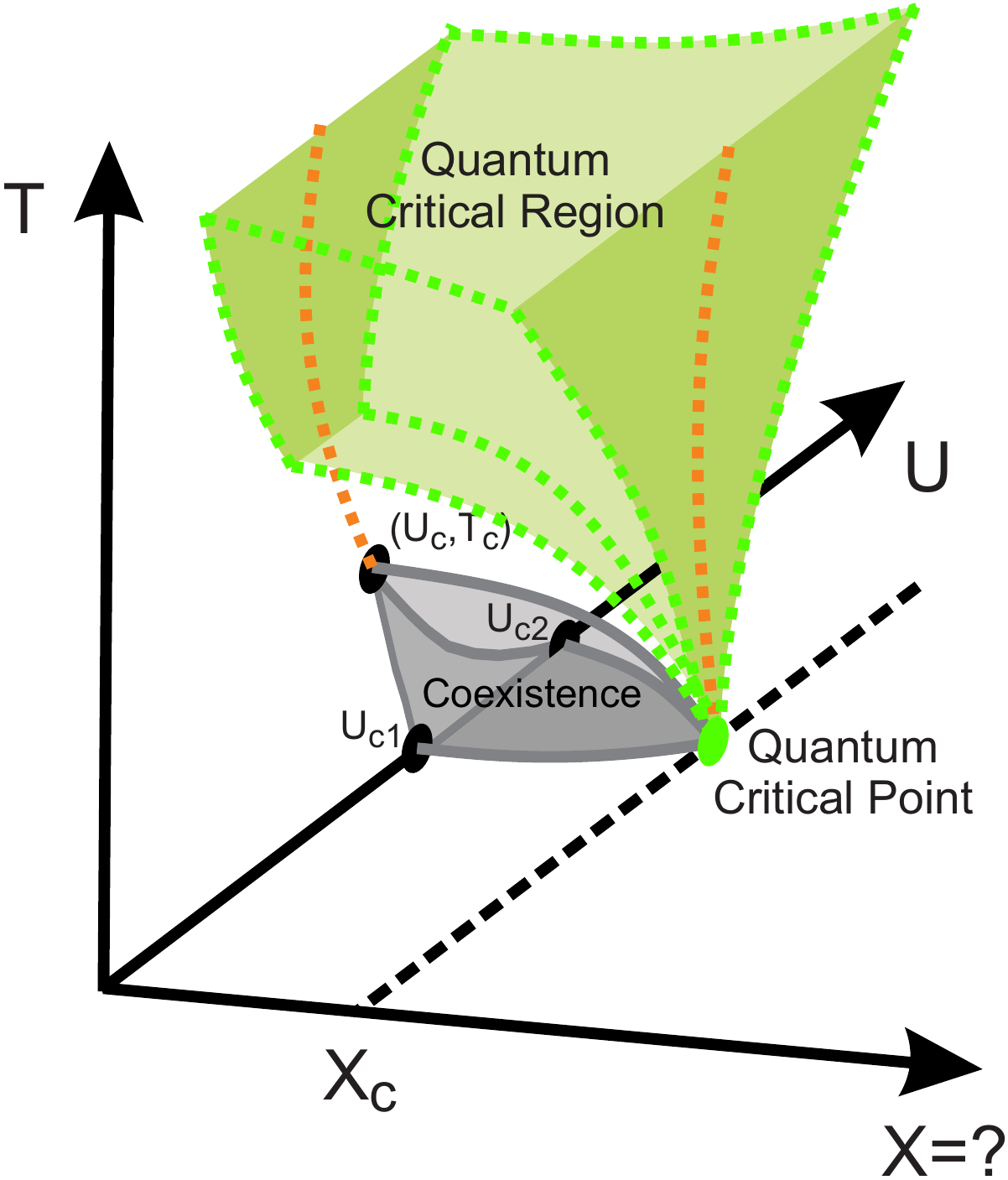}
\caption{ Possible phase diagram of a generalized Hubbard model. The observed scaling (valid in the green region) may be due to a quantum critical
point unreachable by the simple 2-parameter half-filled Hubbard model. An additional, third parameter (here marked with $X$)
could drive $T_c$ to zero at some critical value, and extend
the region of validity of the scaling formula in the $U-T$ plane.
 } \label{QPT_drawing}
\end{figure}

A stringent test of the proposed quantum critical transport scenario would be on the systems with reduced critical temperature $T_c$. Fig.~\ref{QPT_drawing} presents a schematic phase diagram with an additional parameter driving $T_c$ to zero at some critical value $X_c$ and merging $U_{c1}$, $U_{c2}$ and $U_c$ into a single, quantum critical point. If this were the case, the quantum critical region
would extend down to zero temperature. For a simple half-filled Hubbard model, the critical temperature can be reduced, e.g.~ by the disorder\cite{Aguiar2005} or particle-hole asymmetry, but still remains finite. Therefore, other models should be considered, also away from half-filling,\cite{Sordi2009,Amaricci2010} which have significantly reduced coexistence region and where the proposed scaling may give a more direct evidence of the quantum criticality. In some of these models the coexistence region was not even detected, and then the eigenvalue analysis can also be used as an ultimate test for its existence. 
It would be also very interesting to explore a possible quantum critical scaling in the external electric field within the nonlinear $I-V$ regime,\cite{re:Sondhi97} similar as in the experiments on Si MOSFETs.\cite{Kravchenko_finiteE_1996} This seems especially important in the light of the recent discovery of devices displaying 
novel resistive switching in narrow gap Mott insulators.\cite{Guiot2013}
Finally, the concept of the instability line above the quantum critical point, which is based on the thermodynamic analysis, is very general and can be applied to other physical systems (e.g.~interacting spins in an external field) and the scaling analysis can be tested on physical quantities other than the resistivity.

\section{Scaling around the inflection points line}

As stated in the previous section, the curvature $\lambda$ must be directly related to an appropriate relaxation rate of a system perturbed away from the equilibrium, a
quantity that in principle should be possible to measure on any system.
However, it is currently very hard to make such measurements on the Mott systems
and precisely determine the instability line. Our calculations, however, show that it lies just among the crossover lines that separate the bad metal and the Mott insulator, so it might not be necessary
to know its exact position to observe quantum criticality.
In the following, we present a scaling analysis that can be performed around the resistivity inflection points line (or any of the other crossover lines)
to test the scaling hypothesis. As it turns out, the scaling is a robust feature, not particularly sensitive to the choice of $U_c(T)$ as already tested in experiments on various organic Mott systems.\cite{Kanodaprivate}

\begin{figure}[t]
\includegraphics  [width=3.2in]{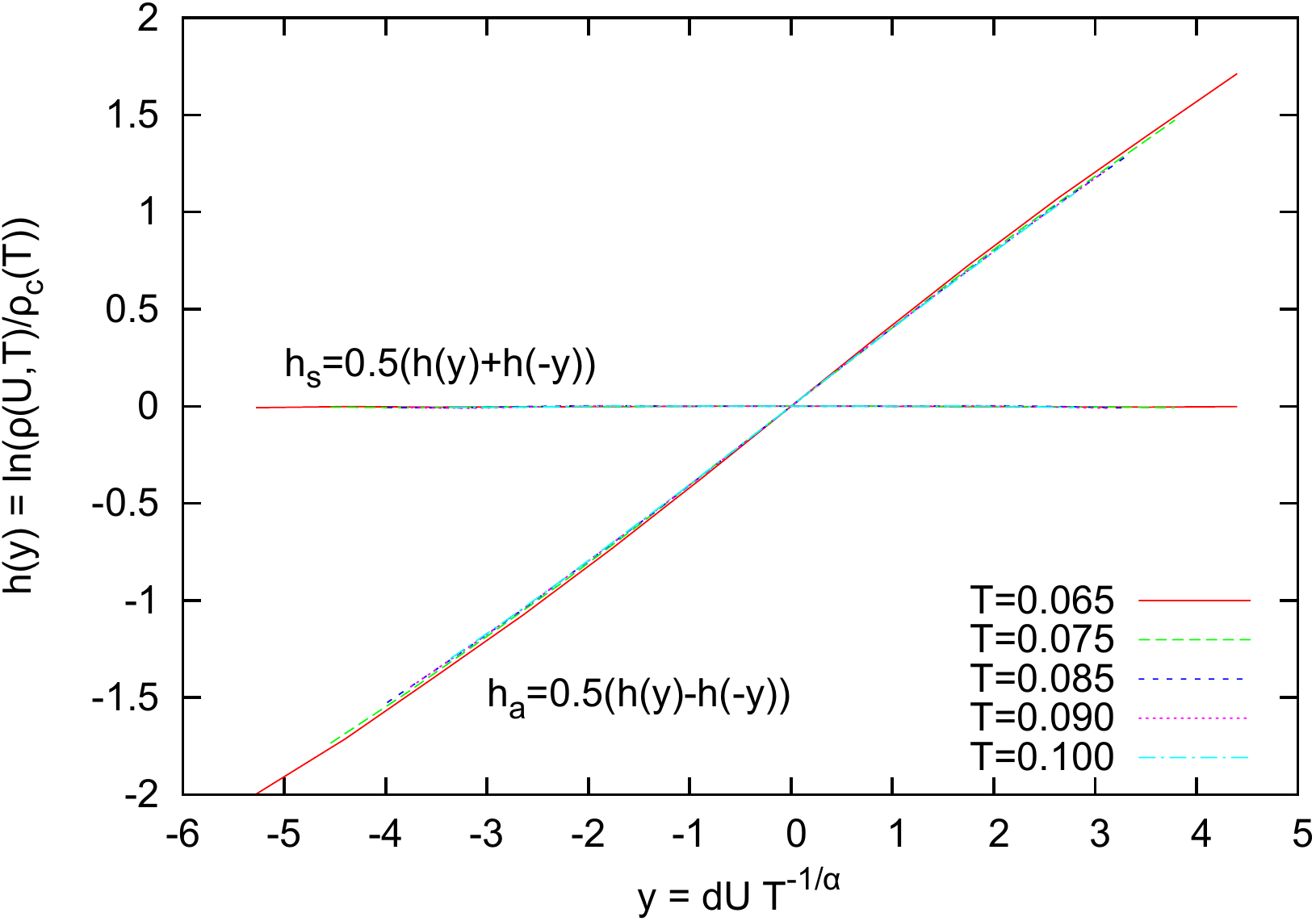}
\caption{ The symmetric and asymmetric part of scaling function, $h_s$ and $h_a$ at various temperatures. The small value of $h_s(y)$ shows that the mirror symmetry of resistivity curves is present. The $h_a(y)$ curves collapse around the inflection-point line which shows that the exponent, $z\nu=0.953$, is well evaluated. Fitting a third order polynomial to $h_a(y)$ in the range where these curves collapse can reveal the exact form of the scaling formula. In our calculations only the
linear term is significant.}
\label{CubicTest}
\end{figure}

We first observe that the resistivity curves display almost a perfect mirror symmetry when plotted on the log-scale (Fig.~\ref{Scaling}(b)). This puts a strong constraint on the functional form of the scaling function $f$ (as we show below) and also indicates that the resistivity curve along the
inflection points line, $\partial \log \rho(U)/ \partial U = 0$, could also serve as the separatrix.
The mirror symmetry requires that
\begin{equation} \label{mirror_f}
  f(y) = 1/f(-y).
\end{equation}
For the above to be satisfied, the function $f$ must be of the form
\begin{equation} \label{mirror_h}
 f(y)=e^{h(y)},
\end{equation}
where $h$ is an antisymmetric function of $y$. It is clear that $f(0)=1$ and therefore $h(0)=0$. $h$ must also be smooth, so it can be represented as a Taylor series with only odd terms
\begin{equation} \label{taylor_h}
 h(y)=ay+by^3+...
\end{equation}
In our calculations, it turns out that only the linear term is significant, and here we show how this can be tested.
First we make a substitution of variables $T/\delta U^{z\nu} \rightarrow \delta UT^{-1/z\nu}$ and then
take the logarithm of both sides of the scaling formula to obtain
\begin{equation} \label{log_rho}
\log \left( \frac{\rho(U_c(T)+\delta U, T)}{\rho(U_c(T), T)} \right) = \log(f(\delta UT^{-1/z\nu})) .
\end{equation}
If the mirror symmetry is satisfied, than
\begin{equation} \label{log_rho2}
\log \left( \frac{\rho(U_c(T)+\delta U, T)}{\rho(U_c(T), T)} \right) = h(\delta UT^{-1/z\nu}) ,
\end{equation}
which means that the precise form of $h(y)$ can be deduced by plotting the left-hand side of the above equation as a function of $y=\delta UT^{-1/z\nu}$ and
than making a fit of a polynomial curve to the data. This is possible because in the region where the scaling formula is valid, all the data points should collapse onto a single curve. To test whether $h(y)$ is truly antisymmetric, it is convenient to first split it into symmetric and antisymmetric parts, $h(y)=h_s(y)+h_a(y)$, where $h_s(y) = \frac{1}{2}\left(h(y)+h(-y)\right)$
and $h_a(y) = \frac{1}{2}\left(h(y)-h(-y)\right)$.
If the resistivity is mirror symmetric, $h_s$ should be 0 and $h_a$ should be equal to $h$.
In Fig.~\ref{CubicTest} we plot these functions around the inflection point line and find $h_s$ to be negligible. Also, it is easily seen that $h(y)$ is purely linear in the region where the data points perfectly collapse on a single curve.

\begin{figure}[t]
\includegraphics  [width=3.2in]{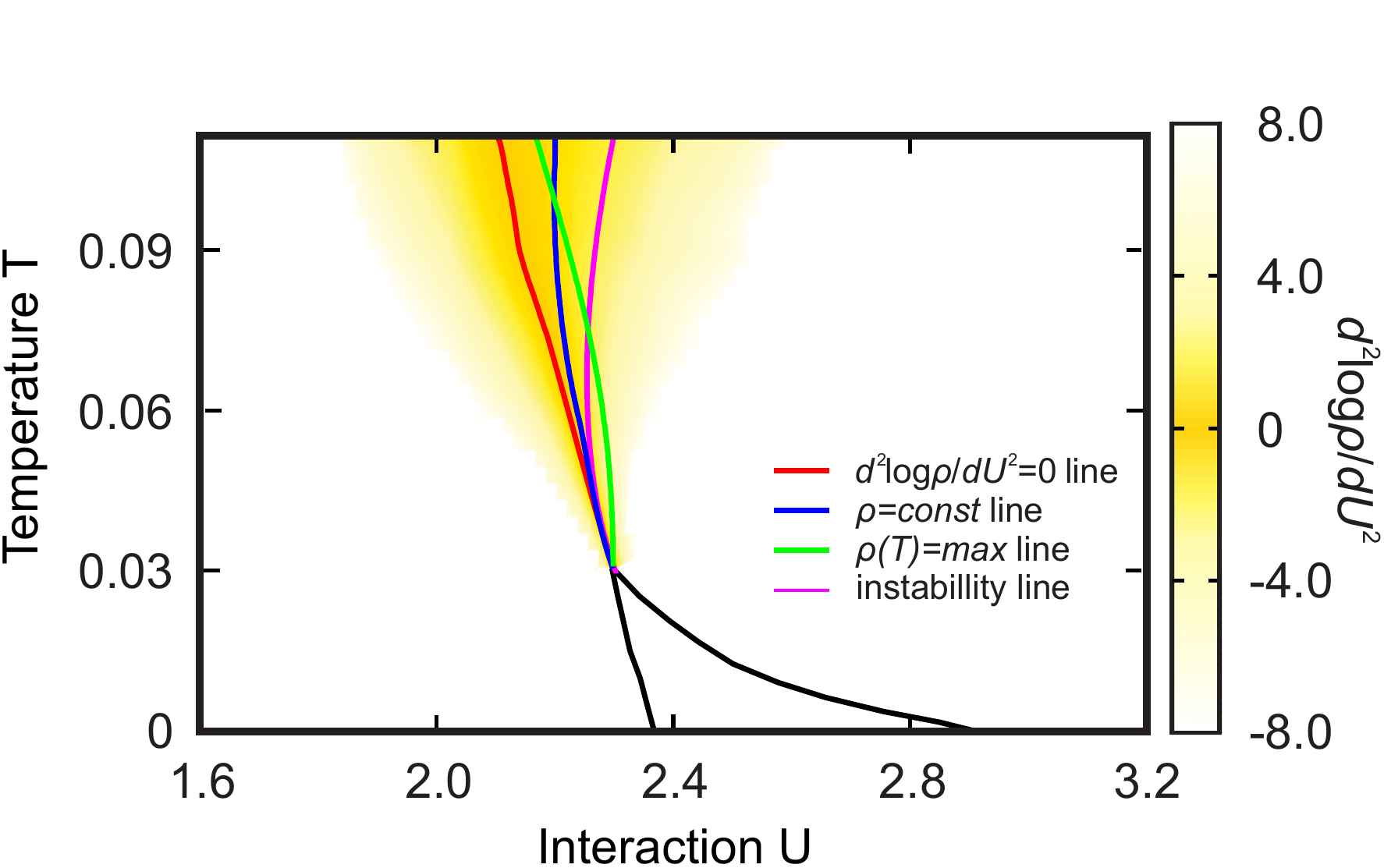}
\caption{ Instability line lies among the other crossover lines.
$\log \rho(U)$ is linear in this crossover region,
which allows for the scaling formula to be valid. }
\label{SecondDer}
\end{figure}

\begin{figure}[t]
\includegraphics  [width=3.2in]{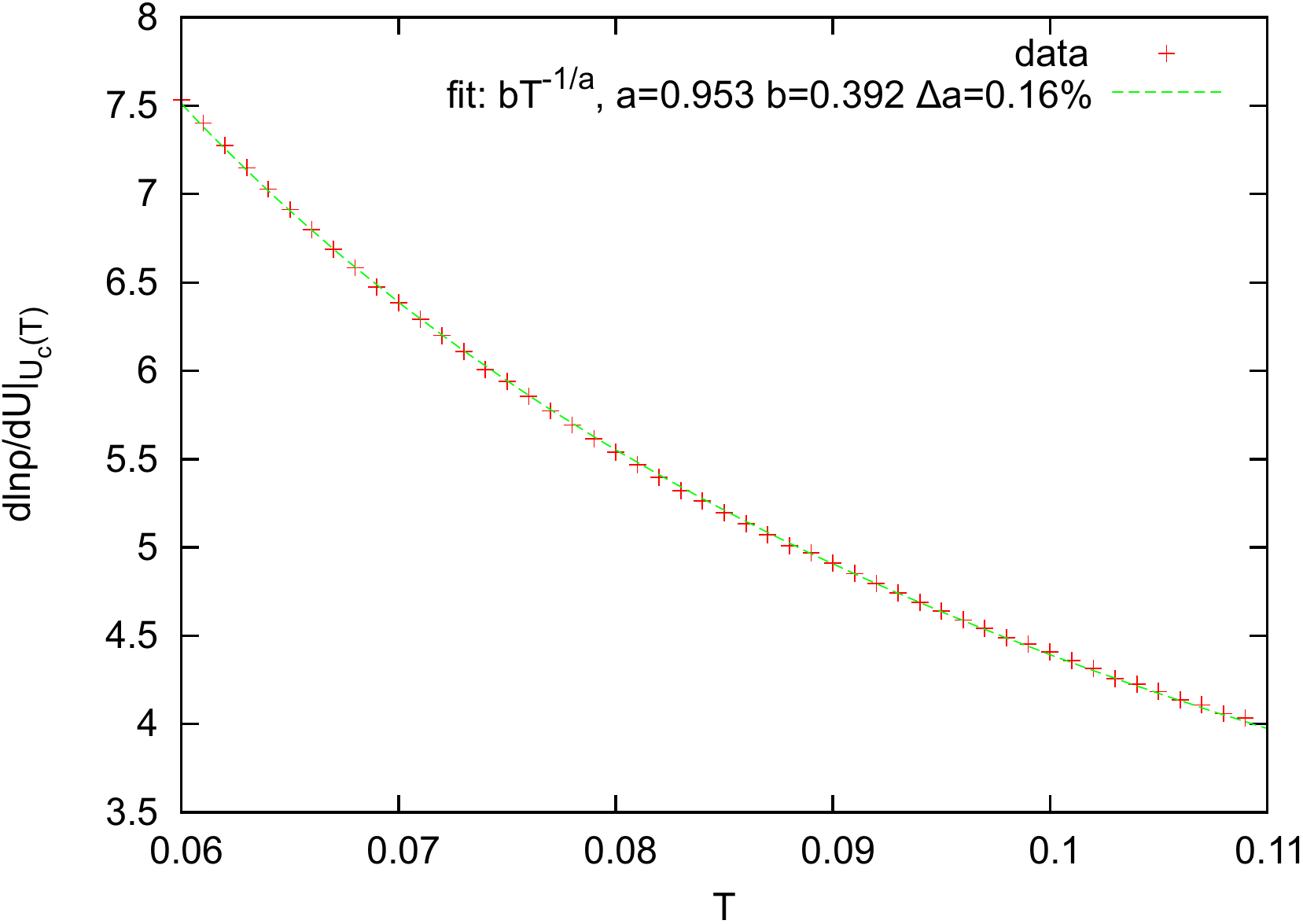}
\caption{ The derivative of resistivity with respect to U ($\left.\partial \rho(U,T)/\partial U \right|_{U_{\mathrm{infl}}}$)
along the inflection-point line. Above roughly $2T_c$, it fits well to a power law curve of exponent $-0.95$.
This can be used to evaluate the value of scaling formula exponent.
At lower temperatures the decrease in resistivity is faster and the behavior deviates from the power-law and the scaling formula fails at temperatures below $2T_c$.
}
\label{Exponent}
\end{figure}

Now it is clear that there are two conditions that $U_c(T)$ has to satisfy for the scaling with mirror symmetry to be possible. First, if we take the partial derivative over $U$ at both sides of the equation, we get
\begin{equation} \label{partial_h}
\frac{\partial \log\rho(U,T)}{\partial U} =  aT^{-\frac{1}{z\nu}} + b\delta U^2 T^{-\frac{3}{z\nu}}+....
\end{equation}\label{taylor}
If $h(y)$ is a linear function, then only the first term in the above equation remains,
which means that the logarithm of resistivity is a linear function of $U$ in the
entire region in which the scaling formula holds. Even if there are higher terms in $h(y)$, the above has to be true at least close to $U_c$ (small $\delta U$), where the linear term is dominant in any case. This imposes a constraint on $U_c(T)$,
such that it has to be in a region where the second derivative of logarithmic resistivity is zero, or at least small,
\begin{equation} \label{condition1}
\frac{\partial^2 \log\rho(U,T)}{\partial U^2} \approx 0.
\end{equation}
This derivative is color-coded in the $(U,T)$ plane in Fig.~\ref{SecondDer}
so that yellow color corresponds to a small absolute value. As it is readily verified, the above condition is not fulfilled anywhere exactly (except precisely at the $\log\rho(U)$ inflection point line by its definition), but all of the crossover lines
lie in the region where this condition is approximately satisfied.
There is an additional requirement for $U_c(T)$ which is not in any way implied by definition of any of the crossover lines.
%The IPT solution suggests that it is best met by the instability line,
%rather than the inflection point line, but from the QMC this is less clear as the lines are very close, and no real distinction in the quality of scaling can be made.
Namely, the first derivative of the logarithmic resistivity has to be decreasing along $U_c(T)$ as a power law of temperature. This can be shown
by taking the limit $\delta U \rightarrow 0$ in Eq.~(\ref{partial_h}),
\begin{equation} \label{condition2}
\left. \frac{\partial \log\rho(U,T)}{\partial U} \right|_{U_c} \propto T^{-\frac{1}{z\nu}} .
\end{equation}
The above holds regardless of the value of the cubic (or any higher) term coefficient.
One can even use this to give a good assessment of the exponent $z\nu$, by fitting such experimental (or theoretical) curve to a power law as shown in Fig.~\ref{Exponent}.
As it is seen here, the derivative Eq.~\ref{condition2} calculated along the inflection point line fits well to a power law curve of exponent $-0.95$, but only above roughly $2T_c$. The same analysis of the IPT results yields a slightly lower value $z\nu = 0.63$.

\begin{figure}[t]
\includegraphics  [width=3.2in]{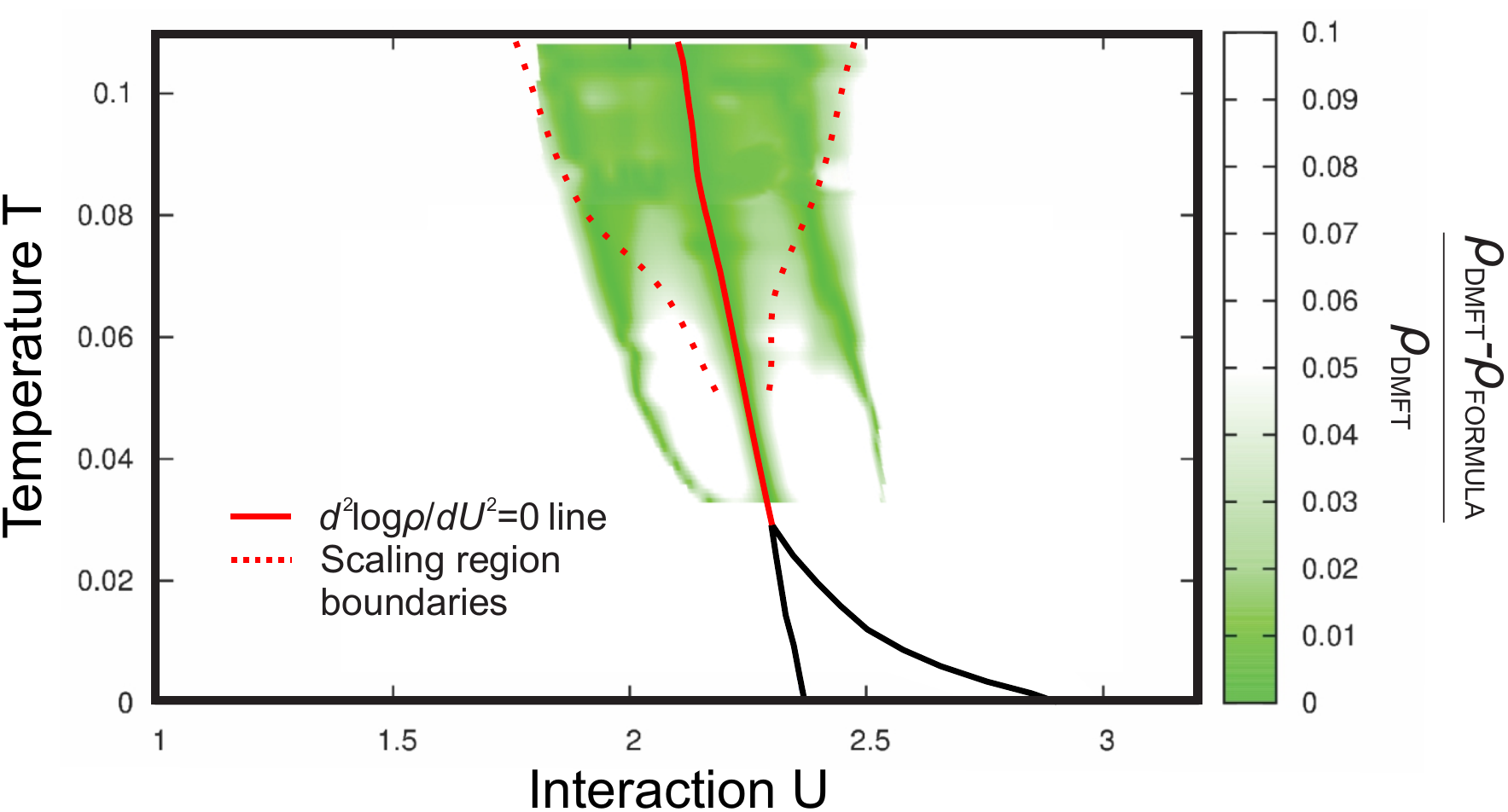}
\caption{ Relative error of the scaling formula color-coded in the $U-T$ plane. The dotted lines are the boundary of the scaling region. Two green filaments below $2T_c$ are where the scaling formula
intersects with the actual DMFT result.}
\label{RelativeError}
\end{figure}

Finally, an estimate of how well the scaling works can be made by comparing the value of resistivity obtained by the scaling formula
and the one measured in experiment or, as it is in our case, calculated from the DMFT solution. In Fig.~\ref{RelativeError} it is shown how the scaling formula works within the $5\%$
error bar in a large region, for the inflection point line. This result is qualitatively the same for the other crossover lines.
It is important to note that in the case of instability line (and all the other crossover lines other than the inflection point line), one is able to improve the quality of scaling by using different exponents $z\nu$ depending on $\mathrm{sign}(\delta U)$, and that way compensate for the lack of exact mirror symmetry.
Also, when only the linear term in $h(y)$ is used, slightly lowering the value of $z\nu$ obtained from the power-law fitting procedure typically broadens the region of validity of such scaling formula.

In conclusion, the $\log\rho(U)$ inflection points line is easily observable in experiment and our calculations show that it lies very close to the instability line.
The analysis presented here, indicates that the quantum critical scaling previously found to hold around the instability line, should also be observable around the inflection point line.
We show that the scaling formula valid around this line displays almost a perfect mirror symmetry of resistivity curves. In general, mirror symmetry, or ``duality'', should not be considered a necessary ingredient for a quantum critical scaling. In fact, we find that the scaling is of better quality around the instability line, altough it is slightly less symmetric.

\begin{figure}[t]
\includegraphics  [width=3.2in]{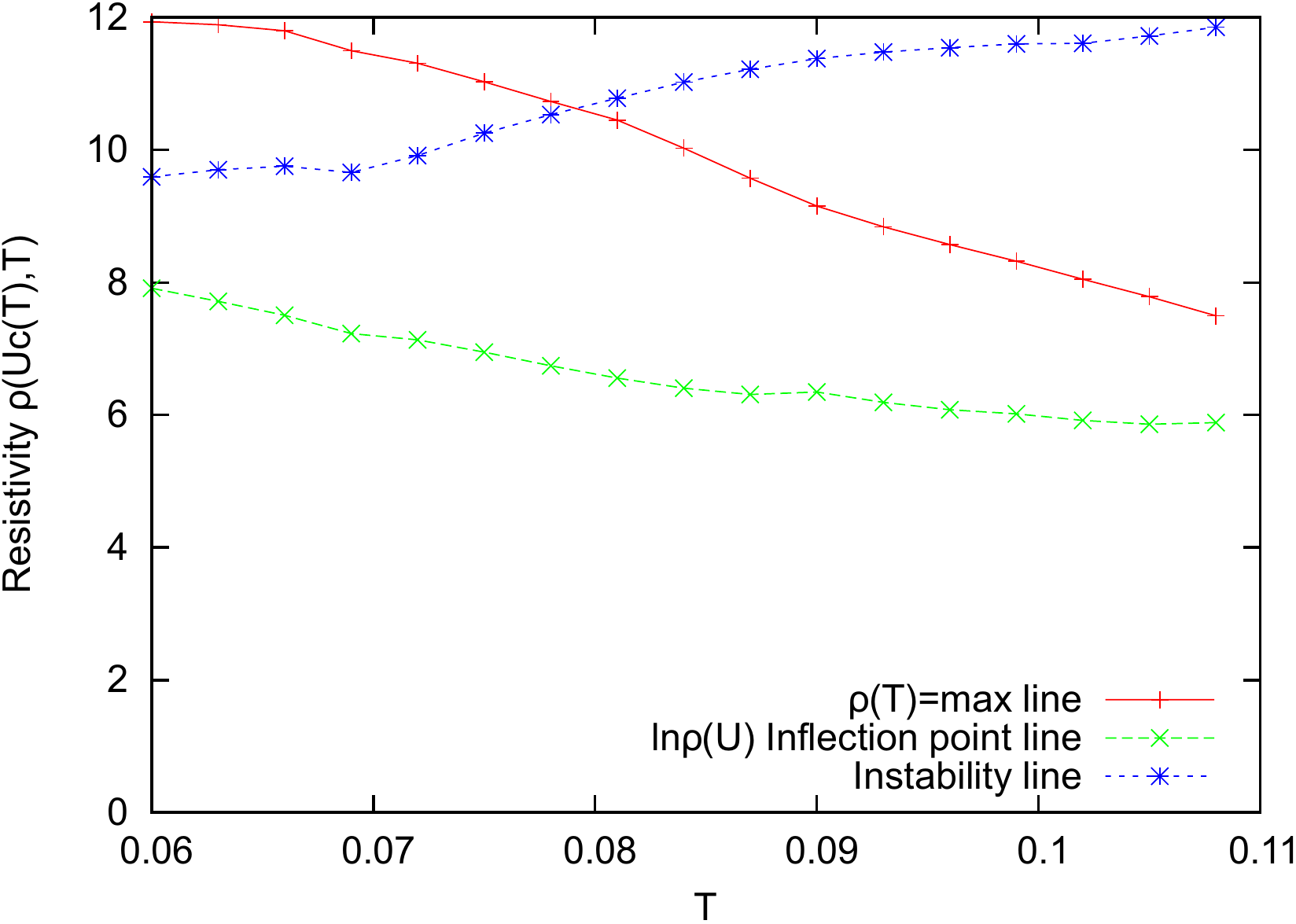}
\caption{ Resistivity (in units of $\rho_{_{Mott}}$) along the crossover lines is weakly dependent on temperature and much larger than the Mott limit. }
\label{ResistivityAlong}
\end{figure}

It is also very important to examine how the resistivity changes along the separatrix and our results are presented in Fig.~\ref{ResistivityAlong}. In this crossover region, the resistivity far exceeds the Mott limit and is only weakly dependent on temperature. We find that along the instability line, the resistivity is roughly a linear, increasing function of $T$. Along the inflection points line and $\rho(T)=max$ lines, the resistivity is slowly decreasing. We note that these results, however, must be model specific. Above the critical end-point, the resistivity is strongly dependent on $U$, and a small change in the shape or position of these lines can cause a significant change in the temperature dependences of resistivity presented in Fig.~\ref{ResistivityAlong}.

\section{Widom lines}

The notion of a crossover line is very general and different physical motivations can be used for its precise definition. The concept of the Widom crossover line is, however, more strict and relies on one fundamental principle.

The Widom line was originally defined in the context of liquid-gas phase transition,\cite{Widom1972} and as the line connecting the maxima of the isobaric specific heat as a function of pressure ($\partial C_p/\partial p=0$), above $T_c$. It was conceived as a logical continuation of the first order phase transition line to supercritical temperatures. $C_p$ is divergent along the first order transition line, which directly causes the maxima in $C_p$ present above the critical temperature.
This concept is easily generalized to include all the lines
that mark features directly caused by non-analyticities due to a phase transition.\cite{Xu_Widomlines2005}
As such, a Widom line can be defined for any quantity that exhibits either a divergence or a discontinuity because of a phase transition, and thus a maximum or an inflection point above $T_c$.

Very recently,\cite{Simeoni2010} in the super-critical region of argon liquid-gas phase diagram, an unexpected non-analyticity has been found in sound velocity dispersion curves, precisely at the Widom line. The authors give a new depth and physical meaning to the concept, by observing that there is no single super-critical fluid phase, and that the Widom line actually separates two regimes
of fluid-like and gas-like dynamical behavior. This finding makes it clear
that the Widom lines should not be exclusively connected with thermodynamics
of the system. The changes in transport that follow certain features in thermodynamic quantities
can also be used for making meaningful and possibly even equivalent definition of the Widom line.
The significance of this concept was recognized once more\cite{Sordi2012,Sordi2013} in the context of hole-doped high-$T_c$ superconductors, where the characteristic temperature $T^*$ of the pseudogap phase is shown to correspond to the Widom line arising above a first-order transition at critical doping.

In the above sense, we emphasize that the quantum critical scaling observed in our model can also be easily connected with the concept of Widom lines, giving them new physical importance in the context of quantum phase transitions. One can immediately recognize that the $\log\rho(U)$ inflection-point line and the instability line both qualify as generalized Widom lines -
they emanate from the the critical end-point, separate regions of metallic and insulating behavior and mark features that are directly caused by non-analyticities due to the phase transition.
The quality of the scaling and the close proximity of these two lines may even
indicate a profound connection between them. As the proposed physical concept may well surpass the scope of the Hubbard model and Mott physics, a definition of the instability line can be very useful. Contrary to the inflection-point line, it is based on a purely thermodynamical quantity, i.e. the free energy, and can be defined for an arbitrary model. It does not require the presence of the finite-temperature critical point (which makes a conceptual difference with the work\cite{Sordi2012,Sordi2013} on hole-doped cuprates) and can be used to introduce the Widom line concept to exclusively zero temperature quantum-phase transitions.

\section{Conclusions}

In this paper we carefully investigated the finite temperature
crossover behavior around the Mott transition, with the goal to
provide both theoretical insight and experimental guidance for the
search of quantum criticality in this regime. To obtain
quantitative and reliable results that allow direct comparison
with experiments, we performed these studies within the framework
of single-site dynamical mean-field theory. From the
conceptual point of view, this approach offers an immediate
advantage - it is physically very clear what kinds of mechanisms
and processes are captured by such a theory, and which are not.
Most importantly, such an approach explicitly excludes all
mechanisms directly or indirectly associated with any ordering
tendencies, in agreement with the physical pictures for the Mott transition introduced by early pioneering ideas of Mott and Anderson.

More specifically, we focused on a single band half-filled Hubbard model, which within DMFT maps to solving a Kondo-Anderson magnetic impurity model in a self-consistently determined bath. The formation of the heavy Fermi liquid on the metallic side of the Mott transition is described as a formation of a Kondo-like singlet in the ground state, similarly as in the early work of Brinkmann and Rice.\cite{Brinkman1970} In contrast to the Brinkmann-Rice theory, the DMFT approach is able to quantitatively and accurately describe the thermal destruction of such a correlated Fermi liquid, and the resulting coherence-incoherence crossover. The possibility to systematically and quantitatively describe this incoherent regime is especially important to properly characterize the high temperature crossover behavior above the coexistence dome, where we obtained clear and precise signatures of quantum critical behavior. Our results show remarkable agreement with several experimental systems,\cite{Kanodaprivate} but future experiments should provide even more precise tests for our predictions. We expect that close enough to the quantum critical point all quantities should display appropriate scaling behaviors. Our work has, so far, focused mostly on the transport properties, and sufficiently detailed results for thermodynamic and other quantities are not available at this time, to permit a scaling analysis. The investigation of  these interesting questions is beyond the scope of the present work, and is left for future studies.

We should mention that ideas closely related to ours have also been discussed in a series of papers by Senthil and collaborators,\cite{Senthil2008a,Senthil2008b,Senthil2012} who also seek a description of Mott quantum criticality unrelated to any ordering phenomena. This approach, however, focuses on capturing the possible effects of gapless "spinon" excitations, which may exist on the insulating side of the Mott transition, but only in presence of sufficient and specific magnetic frustration, preventing the familiar antiferromagnetic order. Because of their gapless nature, they should remain long-lived (e.g. well defined) only at the lowest temperatures, inducing long-range spatial correlations in the proposed spin liquid. The corresponding theory, therefore, focuses on long-distance spatial fluctuations, which as in ordinary critical phenomena, are tackled by appropriate renormalization-group methods. In contrast to our DMFT approach, this theory implicitly disregards the strongly incoherent Kondo-like processes, which may play the dominant role at sufficiently high temperatures.

The key physical question thus remains: What is the crossover temperature $T_{nonlocal}$ below which the nonlocal effects ignored by DMFT become significant? This important question can, in principle, be investigated by computing systematic nonlocal corrections to single-site DMFT, a research direction already investigated by several authors.\cite{Park2008,Liebsch2009,Tanaskovic2011,Georges2011} The recent work already provides some evidence that for a Hubbard model on a square lattice the nonlocal corrections are very small well above the coexistence dome (at $T \gg T_c$)\cite{Park2008} and esentially negligible for frustrated triangular lattice.\cite{Liebsch2009}  On the experimental side, the possible role of nonlocal effects such as spinons can be investigated by systematic studies of a series of materials with varying degrees of magnetic frustration. Such studies are accessible in organic Mott systems,\cite{Kanoda_review2011,Powell_review2011} where $T_c \sim 10-20K$, while the magnetic frustration may be varied using different crystal lattices. In some cases the magnetic ordering is completely suppressed on the insulating side,\cite{Shimizu2003} while in other it remains.\cite{Miyagawa1995} If robust signatures of quantum criticality in transport are observed at $T \gg T_c$ in all of these materials, this finding would provide strong support for the "local quantum criticality" scenario we proposed based on the DMFT approach.
%%%%%%%%%%%%%%%%%%%%%%%%%%%%%%%%%%%%%%%%%%%%%%%%%%%%%%%%%%%%%%%%%%%%%%%%%%%%%
%%%%%%%%%%%%%%%%%%%%%%%%%%%%%%%%%%%%%%%%%%%%%%%%%%%%%%%%%%%%%%%%%%%%%%%%%%%%%

\begin{acknowledgments}
We thank A. Georges, K. Kanoda, G. Kotliar, M. Rozenberg, G. Sordi and J. M. Tremblay for useful discussions. The authors thank K. Haule for the usage of his CTQMC code.
J.V. and D.T. acknowledge support from the Serbian
Ministry of Education and Science under project No.~ON171017. V.D.
was supported by the National High Magnetic Field Laboratory and
the NSF Grant No.~DMR-1005751, and H.T. by the DOE BES CMCSN grant
DE-AC02-98CH10886. Numerical simulations were run on the AEGIS
e-Infrastructure, supported in part by FP7 projects EGI-InSPIRE,
PRACE-1IP and HP-SEE.
\end{acknowledgments}

\appendix

\bibliographystyle{apsrev}
%\bibliography{vucicevic_QWL}

\end{document}